\documentstyle[12pt]{article}   % defines the document style

\large
\newcommand{\beq}{\begin{equation}}
\newcommand{\eeq}{\end{equation}}

\makeatletter
\@addtoreset{equation}{section}
\makeatother

\newtheorem{Theorem}{Theorem}[section]

\newtheorem{Lemma}{Lemma}[section]

\def\be{\begin{equation}}
\def\ee{\end{equation}}
\def\ba{\begin{eqnarray}}
\def\ea{\end{eqnarray}}

\def\agb{{\overline {{\cal A}/{\cal G}}}}

\def\Comp{{\mathchoice
{\setbox0=\hbox{$\displaystyle\rm C$}\hbox{\hbox to0pt
{\kern0.4\wd0\vrule height0.9\ht0\hss}\box0}}
{\setbox0=\hbox{$\textstyle\rm C$}\hbox{\hbox to0pt
{\kern0.4\wd0\vrule height0.9\ht0\hss}\box0}}
{\setbox0=\hbox{$\scriptstyle\rm C$}\hbox{\hbox to0pt
{\kern0.4\wd0\vrule height0.9\ht0\hss}\box0}}
{\setbox0=\hbox{$\scriptscriptstyle\rm C$}\hbox{\hbox to0pt
{\kern0.4\wd0\vrule height0.9\ht0\hss}\box0}}}}
\def\Co{{\mathchoice
{\setbox0=\hbox{$\displaystyle\rm C$}\hbox{\hbox to0pt
{\kern0.4\wd0\vrule height0.9\ht0\hss}\box0}}
{\setbox0=\hbox{$\textstyle\rm C$}\hbox{\hbox to0pt
{\kern0.4\wd0\vrule height0.9\ht0\hss}\box0}}
{\setbox0=\hbox{$\scriptstyle\rm C$}\hbox{\hbox to0pt
{\kern0.4\wd0\vrule height0.9\ht0\hss}\box0}}
{\setbox0=\hbox{$\scriptscriptstyle\rm C$}\hbox{\hbox to0pt
{\kern0.4\wd0\vrule height0.9\ht0\hss}\box0}}}}
\def\Rl{{\mathchoice
{\setbox0=\hbox{$\displaystyle\rm R$}\hbox{\hbox to0pt
{\kern0.4\wd0\vrule height0.9\ht0\hss}\box0}}
{\setbox0=\hbox{$\textstyle\rm R$}\hbox{\hbox to0pt
{\kern0.4\wd0\vrule height0.9\ht0\hss}\box0}}
{\setbox0=\hbox{$\scriptstyle\rm R$}\hbox{\hbox to0pt
{\kern0.4\wd0\vrule height0.9\ht0\hss}\box0}}
{\setbox0=\hbox{$\scriptscriptstyle\rm R$}\hbox{\hbox to0pt
{\kern0.4\wd0\vrule height0.9\ht0\hss}\box0}}}}

\def\n{\nu}

\title{Quantum Spin Dynamics (QSD)}
\author{T. Thiemann\thanks{thiemann@math.harvard.edu} \\
       Physics Department, Harvard University, \\
       Cambridge, MA 02138, USA}
\date{{\small Preprint HUTMP-96/B-351}}

\begin{document}

\maketitle

\begin{abstract}
An anomaly-free operator corresponding to the
Wheeler-DeWitt constraint of Lorentzian, four-dimensional, canonical,
non-perturbative vacuum gravity is constructed in the continuum. This 
operator is entirely free of factor ordering singularities and can be
defined in symmetric and non-symmetric form.

We work in the real connection representation and obtain a well-defined
quantum theory. We compute the complete solution to the Quantum Einstein
Equations for the non-symmetric version of the operator and a physical 
inner product thereon.

The action of the Wheeler-DeWitt constraint on spin-network states is by 
annihilating, creating and rerouting the quanta of angular momentum 
associated with the edges of the underlying graph while the ADM-energy
is essentially diagonalized by the spin-network states. We argue that 
the spin-network representation is the ``non-linear Fock representation" of 
quantum gravity, thus justifying the term ``Quantum Spin Dynamics (QSD)". 
\end{abstract}

\section{Introduction}

Attempts at defining an operator corresponding to the Wheeler-DeWitt
constraint \cite{1} of Lorentzian, four-dimensional canonical vacuum gravity 
have been made first in the framework of the metric (or ADM) variables
(see, for instance, \cite{2}). The formulation of the theory seemed
hopelessly difficult because of the complicated, non-polynomial algebraic 
form of the Wheeler-DeWitt constraint and it was therefore thought to be 
mandatory to first obtain a polynomial formulation of the theory to 
complete the programme.

The celebrated observation due to Ashtekar \cite{3} is that the canonical 
constraints
of general relativity can indeed be cast into polynomial form if one 
performs a 
certain {\em complex} canonical transformation on the gravitational phase 
space.
This opened, for the first time, the hope that one can actually rigorously
define the quantum Hamiltonian constraint (or Wheeler-DeWitt equation).\\ 
The major
roadblock after this was the difficult reality structure of the theory so
obtained : general relativity, when written in Ashtekar's variables, is a 
dynamical theory of complex-valued connections for the 
non-compact gauge group $SL(2,\Co)$, however, standard mathematical 
constructions and techniques usually used for Yang-Mills theory apply only
if the gauge group is compact. 

A solution to this problem has been recently proposed in form of a phase 
space Wick rotation transform \cite{4} (see also \cite{5}) which 
enables one to formulate general relativity as a $SU(2)$ gauge theory while
keeping the polynomial algebraic form of Wheeler-DeWitt constraint and 
incorporating the correct reality conditions into the quantum theory.
This theory can be, not surprisingly, recognized as Euclidean 
four-dimensional gravity as written in terms of {\em real valued} 
Ashtekar variables \cite{6}.
This then opens access to the recently developed, powerful calculus on the 
space of (generalized) connections modulo gauge transformations for compact
gauge groups \cite{7,8,9,10,11}. This calculus provides a rigorous 
kinematical framework by means of which constraint operators can be
regularized in a well-defined and unambiguous manner. In particular,
this framework has already been successfully employed to arrive at the 
general solution of the Gauss and Diffeomorphism constraints \cite{18}.

In order to complete the programme one still needed to construct a 
rigorously defined operator corresponding to the classical generator of the 
transform. This seems a hopeless problem to solve as the classical generator
is a non-polynomial, not even analytic function of the phase space variables.

There is, however, an even more severe problem : the Hamiltonian constraint
of Euclidean or Lorentzian gravity as defined by Ashtekar is a density of
weight two. Namely, in order to obtain a polynomial form of those constraints
one needs to rescale the constraint functional and to absorb a factor of 
$1/\sqrt{\det(q)}$ into the Lagrange 
multiplier (the lapse function; here $q=(q_{ab})$ is the intrinsic metric
of an initial data hypersurface). On general grounds, an operator 
corresponding to a classical function on the phase space with density weight
different from one needs to be renormalized. While this poses no, a priori,
problems for, say, Yang-Mills theories in a classical background geometry,
such a procedure is unacceptable for quantum gravity since a renormalization
unavoidably introduces a length scale and thus breaks diffeomorphism
invariance.

A solution to this problem was first suggested in \cite{12} and is 
currently reconsidered in \cite{13} : to produce an operator of density 
weight one one takes the square root of the rescaled Euclidean Hamiltonian 
constraint. 
While classically this does not alter the theory and while this seems
to be a necessary step to do in quantum theory in order to keep 
diffeomorphism invariance,
there remain problems that have to do with taking the square root of an 
infinite number of non-self adjoint, non-positive, non-commuting operators.

In the present article we suggest a technique to solve both problems, namely 
the complicated reality
structure of the Lorentzian theory as well as the problem connected with the
density weight, in one stroke : 
what was thought to create problems in the quantization process 
is precisely the reason for why we are able to find a finite operator :
the factor $1/\sqrt{\det(q)}$ is {\em needed}.\\
\begin{itemize}
%$\bullet$ 
\item We show that it is indeed possible to construct a finite, 
(symmetric) operator corresponding to the original, non-rescaled, Lorentzian
Wheeler-DeWitt constraint whose quantum constraint algebra is 
non-anomalous. Since the original Wheeler-DeWitt constraint has density 
weight one, no renormalization is necessary.
%$\bullet$ 
\item Amazingly, the resulting operator is not messy at all and the
problem of finding exact solutions to the quantum constraint is 
conceivable.
%$\bullet$ 
\item We always work with real-valued Ashtekar variables, the reality 
structure of the theory is very simple, the complex-valued Ashtekar 
variables are never introduced.
%$\bullet$ 
\item
In the construction of the Wheeler-DeWitt operator the 
{\em unrescaled} Euclidean Hamiltonian operator as well as the generator
of the Wick transform arise in a natural way so that both operators 
are also constructed rigorously as a side result. This is important
since the Wick rotation transform simplifies the problem of finding solutions
to the quantum constraint.
%$\bullet$ 
\item
Using the same technique one can give rigorous meaning to a 
whole bunch of other operators in a representation where the metric is not 
diagonal, including but not exhausting a) the operator corresponding to
the length of a curve \cite{14}, b) the generators of the Poincar\'e 
group for asymptotically flat topologies \cite{14a} and c) matter 
contributions to the 
Hamiltonian constraint \cite{15}. It might be that it is in this sense
that quantum gravity arises as the ``natural regulator of the matter
field theories". By this we mean the following : in the canonical 
quantization programme a natural regularization procedure is point 
splitting. With the exception of spinorial matter, all matter Hamiltonians
of, say, the standard model,
are quadratic in the momenta which means that they display a density weight
of {\em two} when neglecting the gravitational interaction while they 
have density weight {\em one} when taking gravity into account. When 
removing the regulator the density weight shows up in the form of a product of 
delta distributions evaluated at the same point which is singular. On 
the other hand, with gravity the singularity is removed, in the limit
one arrives at a well-defined operator-valued distribution.
\end{itemize}
The article is organized as follows :

After fixing the notation and explaining the main idea we construct 
first the unrescaled Euclidean Hamiltonian constraint operator. We motivate
our choices involved in the regularization step. The freedom in our
choices is severely restricted by the requirement that the resulting
operator be diffeomorphism-covariantly defined and non-anomalous.
We provide a solution to both requirements.
If one adds the requirement that the operator be at least symmetric 
(preferrably, it should possess a self-adjoint extension) then the 
regularization involves 
an additional structure. We will stick with a non-symmetric operator in 
the main text and provide a symmetric operator in \cite{0}
which is non-anomalous and diffeomorphism-covariantly defined as 
well.

Next we construct operators corresponding to the generator of the Wick 
rotation transform and finally the Lorentzian Wheeler-DeWitt operator.

For the non-symmetric operator we are able to find the complete kernel
of the Wheeler-DeWitt constraint operator in \cite{0}. The physical Hilbert
space turns out to be the one already given in \cite{18}.

For the symmetric operator on the other hand we do not have the complete
solution yet, although solutions can be easily computed by a case by case 
analysis. 
In principle, since, expectedly, on diffeomorphism invariant states the 
constraint algebra is Abelian we are able to 
find solutions to the constraint as well as constructing a physical 
inner product by the group averaging method \cite{16,17}. 
We also comment on finding observables along the line of argument in
\cite{18}. A lot of the mathematical problems of quantum gravity can be 
solved, 
on so-called cylindrical subspaces, with well-known Hilbert space techniques 
familiar from quantum mechanics.

We have complete control over the space of solutions to both 
versions of 
the constraint and the intuitive picture that arises is the following :
The Hamiltonian constraint acts by annihilating, creating and re-routing 
the quanta of angular momentum (with which the graphs of so-called 
spin-network \cite{19,20,21} states are ``coloured") in units 
of $\pm\hbar,\pm\hbar/2,0$. On the other hand, linear combinations of 
such states diagonalize the ADM energy operator very much in the same 
way as Fock states diagonalize the Maxwell Hamiltonian, the role of the 
occupation number being played by the spins of the spin-network 
state. Thus, the spin-network representation is the ``non-linear Fock
representation for quantum gravity" and this motivates to call the quantum 
theory we obtain ``Quantum Spin Dynamics (QSD)" in analogy with QED or QCD.

\section{Notation and the main idea}

Let the triad on the spacelike, smooth, hypersurface $\Sigma$ be 
denoted by $e_a^i$,
where $a,b,c,...$ are tensorial and $i,j,k,...$ are $SU(2)$ indices.
The relation with the intrinsic metric is given by $q_{ab}=e_a^i e_b^j
\delta_{ij}$. It follows that $\det(q):=\det((q_{ab}))=[\det((e_a^i))]^2\ge 
0$. The
densitized triad is then defined by $E^a_i:=\det((e_b^j))e^a_i$ where
$e^a_i$ is the inverse of $e_a^i$. We also need the field $K_a^i=e K_{ab} 
e^b_i,\; e=\mbox{sgn}(\det((e_b^j))),$ where $K_{ab}$ is the extrinsic 
curvature of $\Sigma$. It turns
out that the pair $(K_a^i,E^a_i)$ is a canonical one, that is, these
variables obey
canonical brackets $\{K_a^i(x),E^b_j(y)\}=\kappa\delta^{(3)}(x,y)\delta_a^b
\delta_j^i$ where $\kappa$ is Newton's constant.\\
Let the spin-connection (which annihilates the triad) be denoted by
$\Gamma_a^i$. Then one can show that $(A_a^i:=\Gamma_a^i+K_a^i,E^a_i)$
is a canonical pair\footnote{If we had chosen 
$E^a_i=\sqrt{\det(q)}e^a_i$ instead of $E^a_i=\det(e)e^a_i$ then this
pair is not canonical, a fact often overlooked in the literature 
\cite{3}. Roughly speaking, we would spoil the integrability of the spin
connection.} on 
the phase space of Lorentzian gravity subject
to the $SU(2)$ Gauss constraint, the diffeomorphism constraint and 
the Wheeler-DeWitt constraint (neglecting a term proportional to the
Gauss constraint)
\be \label{1}
H:=\sqrt{\det(q)}[K_{ab}K^{ab}-(K_a^a)^2-R]
=\frac{1}{\sqrt{\det(q)}}\mbox{tr}((F_{ab}-2R_{ab})[E^a,E^b])
\ee
where $F_{ab}$ and $R_{ab}$ respectively are the curvatures of the
$SU(2)$ connection $A_a^i$ and the triad $e_a^i$ respectively.\\
What has been gained by reformulating canonical gravity as a dynamical
theory of $SU(2)$ connections is the following : if, as we do in the sequel, 
one makes the  assumption that there exists a phase for quantum gravity in 
which the
excitations of the gravitational field can be probed by loops rather than,
say, test functions of rapid decrease, then one has access to 
a powerful calculus on the space of (generalized) connections modulo
gauge transformations $\agb$ and, in particular, there is a natural
choice of a diffeomorphism invariant, faithful measure $\mu_0$ thereon which 
equips us with a Hilbert space ${\cal H}:=L_2(\agb,d\mu_0)$, appropriate 
for a representation in which $A$ is diagonal. Moreover, Gauss and 
diffeomorphism
constraints can be solved (see \cite{9} and references therein for
an introduction to these concepts).\\
The remaining step then is to give a rigorously defined quantum operator
corresponding to the Wheeler-DeWitt constraint and to project the scalar
product on its kernel. \\
Let us explain elements of the underlying kinematical framework \cite{18} :\\
We begin by explaining the notion of a ``cylindrical function" on $\agb$.
In brief terms, gauge invariant cylindrical functions on the space of 
(generalized) $SU(2)$ connections
are just finite linear combinations of traces of the holonomy around closed
analytic\footnote{The extension of the framework to the smooth category 
is possible but not entirely straightforward \cite{BaSa}. For simplicity 
we restrict ourselves to the analytic category.} loops in 
$\Sigma$. Each such function therefore may equally well be labelled
by the closed, piecewise analytic graph $\gamma$ consisting of the union of 
all loops involved in that linear combination. 
Such a graph consists of a finite number of edges $e_1,..,e_n$ and vertices 
$v_1,..,v_m$. So, a function cylindrical with respect to a graph $\gamma$
typically looks like
$f(A)=f_\gamma(h_{e_1}(A),..,h_{e_n}(A))$, where $h_e(A)$ is the holonomy
along $e$ for the connection $A$ and $f_\gamma$ is a 
complex-valued function on $SU(2)^n$ such that $f(A)$ is gauge-invariant. 
The functions cylindrical with 
respect to a graph that are $n$-times differentiable 
with respect to the standard differentiable structure on $SU(2)^m$ for 
some $m$ will be denoted by
$\mbox{Cyl}_\gamma^n(\agb)$ and $\mbox{Cyl}^n(\agb):=\cup_\gamma 
\mbox{Cyl}_\gamma^n(\agb)$. 
One and the same cylindrical function can be represented 
on different graphs leading to cylindrically equivalent representants
of that function. It is understood in the above union that such kind of 
functions are identified.\\
The measure $\mu_0$ referred to above is 
entirely characterized by its cylindrical projections defined by 
\[ \int_\agb d\mu_0(A) f(A)=\int_\agb d\mu_{0,\gamma}(A) 
f_\gamma(\{h_{e_i}(A)\})=\int_{SU(2)^n} d^n\mu_H(g_1,..,g_n) 
f_\gamma(g_1,..,g_n)\;. \]\\ 
An orthonormal basis on $\cal H$ is given by the so-called spin-network 
states \cite{19,20,21} : given a graph $\gamma$, ``colour" each of its
edges $e$ with a non-trivial irreducible representation $\pi_{j_e}$
of $SU(2)$, that is, $j_e$ is the spin associated with $e$. With each 
vertex we associate a contraction matrix $c_v$ which contracts all the 
matrices $\pi_{j_e}(h_e)$ for $e$ incident at $v$ in a gauge-invariant way.
In this paper we will denote a spin-network state by 
$T_{\gamma,\vec{j},\vec{c}}$
if we wish to stress the dependence on $\gamma,\vec{j}=(j_e),\vec{c}=(c_v)$ 
where the vectors $\vec{j},\vec{c}$ have indices corresponding to the 
edges and vertices of $\gamma$ respectively.\\
Consider the set of smooth cylindrical functions 
$\Phi:=\mbox{Cyl}^\infty(\agb)$ which can be shown to be dense in $\cal H$.
By a distribution $\psi\in\Phi'$ on $\Phi$ we mean a generalized function on
$\agb$ such that for any $\phi\in\Phi$ the number $\psi[\phi]:=\int_\agb
d\mu_0 \overline{\psi}\phi<\infty$ is finite. It turns out that the 
solutions of the diffeomorphism constraint are elements of $\Phi'$ 
\cite{18}.\\
Suppose we have a quantization $\hat{H}(N)$ of the Hamiltonian constraint 
operator, that is, it is densely defined on $\cal H$ and its classical
limit reduces to $H(N)$, where $N$ is the lapse function. Then its 
adjoint $\hat{H}(N)^\dagger$ also has 
the same classical limit $H(N)$ (a reordering of terms gives only higher 
orders in $\hbar$) because $H(N)$ is real-valued. Therefore, if $\hat{H}(N)$
is not self-adjoint, then it is an option of whether we impose 
$\hat{H}(N)\psi=0$ or $\hat{H}(N)^\dagger\psi=0$. 
In both cases, the solution $\psi$ is typically not an $L_2$ function any 
longer but a distribution, that is, in our case an element of $\Phi'$.
Thus, given the framework of generalized eigenvectors \cite{18} and the 
associated triple $\Phi\subset{\cal H}\subset\Phi'$ we choose to find 
the generalized eigenvectors with eigenvalue $0$ corresponding to the 
kernel of the 
(self-adjoint) Hamiltonian constraint operator $\hat{H}(N)^\dagger$ as 
follows : Let $\psi\in\Phi'$ be a distribution.
We say that $\psi$ is in the kernel of $\hat{H}(N)^\dagger$ whenever
$(\hat{H}(N)^\dagger\psi)(f):=\psi(\hat{H}(N)f)=0$ for each lapse and each 
cylindrical function $f$ in the (dense) domain of $\hat{H}(N)$. 
Note that we cannot require that
$\psi$ is diffeomorphism invariant if we impose the condition
$\hat{H}(N)^\dagger\psi=0$. This is because 
the Hamiltonian constraint does not leave the subspace of $\Phi'$, 
corresponding to diffeomorphism invariant elements, invariant so that one
would be forced to
solve the Hamiltonian constraint {\em before} the diffeomorphism 
constraint. On the other hand, we will see that if $\psi$ is 
diffeomorphism invariant and $\hat{H}(N)$ is diffeomorphism covariantly 
defined, then solving $\psi(\hat{H}(N)f)=0$ is meaningful.\\ \\
After this preparation we are now ready to explain the main idea of our 
approach.\\
Suppose that we can give meaning, in a representation in which $A$ is 
diagonal, to two operators corresponding to\\
1) The total volume of $\Sigma$ given by\footnote{In case that 
$\Sigma$ is not compact, as we will need only the variation of $V$ in the 
sequel, (\ref{2}) is to be understood in the following way : consider
any nested one-parameter family of compact manifolds 
$\Sigma_r\subset\Sigma_{r'},\;\forall\;0\le r<r'<\infty$ such that 
$\lim_{r\to\infty}\Sigma_r=\Sigma$. Then $\delta V(\Sigma):=\lim_{r\to\infty}
\delta V(\Sigma_r)$ and this is well-defined.}
\be \label{2}
V:=V(\Sigma):=\int_\Sigma d^3x \sqrt{|\det(q)|} \mbox{ and}
\ee 
2) the integrated trace of the (densitized) extrinsic curvature of 
$\Sigma$
\be \label{3}
K:=\int_\Sigma d^3x \sqrt{\det(q)}K_{ab} q^{ab}=
\int_\Sigma d^3x K_a^i E^a_i \;. \ee
This means that their quantizations $\hat{V},\hat{K}$ are densely defined
on suitable subspaces of cylindrical functions (in case that we wish 
to obtain a symmetric operator we will also require that they are 
self-adjoint on ${\cal H}$).
The motivation for these two assumptions comes from considering the following
two {\em key} identities
\ba \label{4}
&&\frac{[E^a,E^b]^i}{\sqrt{\det(q)}}
=\epsilon^{abc}e e_c^i(x)
=2\epsilon^{abc}\frac{\delta V}{\delta E^c_i(x)}
=2\epsilon^{abc}\{\frac{A_c^i}{\kappa},V\}\mbox{ and }\\
&& \label{5}
K_a^i=\frac{\delta K}{\delta E^a_i}=\{\frac{A_a^i}{\kappa},K\} \;.
\ea
The last equality relies on the observation that $\{\Gamma_a^i,K\}=0$
and it is this identity underlying the ideas developed in \cite{4}.
The importance of these identities becomes clear when we get rid of the 
complicated 
curvature term $R_{ab}$ involved in (\ref{1}) in favour of $K$. We have
\ba \label{6}
&& H+H^E=\frac{2}{\sqrt{\det(q)}}\mbox{tr}([K_a,K_b][E^a,E^b]) 
=\frac{2}{\sqrt{\det(q)}}\mbox{tr}([\{\frac{A_a}{\kappa},K\},
\{\frac{A_b}{\kappa},K\}][E^a,E^b])
\nonumber\\
&& =\frac{4}{\kappa^3}\epsilon^{abc}\mbox{tr}([\{A_a,K\},\{A_b,K\}]\{A_c,V\})
=\frac{8}{\kappa^3}\epsilon^{abc}\mbox{tr}(\{A_a,K\}\{A_b,K\}\{A_c,V\})
\ea
where we have introduced the (unrescaled) Euclidean Hamiltonian constraint
\be \label{7}
H^E:=\frac{1}{\sqrt{\det(q)}}\mbox{tr}(F_{ab}[E^a,E^b])=
\frac{2}{\kappa}\epsilon^{abc}\mbox{tr}(F_{ab}\{A_c,V\}) \;.
\ee
So what we have achieved is to hide the non-polynomiality of the theory as 
determined by $1/\sqrt{\det(q)}$ in a Poisson bracket. Classically this 
is not helpful at all (except, possibly, for Hamilton-Jacobi methods or
semi-classical approximations), however, we will show that in the quantum 
theory it {\em is} of advantage. Namely, it is now clear where we are 
driving at : in any regularization of 
the Wheeler-DeWitt constraint operator we will have to approximate the
connection $A_a$ and the curvature $F_{ab}$ respectively by 
cylindrical functions given by the holonomies 
$h$ along some edges and closed loops respectively. Now, one obvious 
quantization
of (\ref{6}) would be to replace $V,K$ by $\hat{V},\hat{K}$ and the Poisson
brackets $\{.,.\}$ by $[.,.]/(i\hbar)$. It then follows, given our
assumption, that this quantization of (\ref{6}) has a chance to result in
a finite operator on cylindrical functions since no singular terms appear
when computing the bracket and we would have managed to produce a 
densely defined operator.\\
Is it then true that there exist quantizations of $V,K$ meeting our 
assumption ? The answer is, surprisingly, in the affirmative :\\
First, it is a fact that there has already been constructed a well-defined,
self-adjoint operator $\hat{V}$ on $\cal H$ corresponding to $V$ 
\cite{22,23} whose action on 
cylindrical functions is perfectly finite : (we follow \cite{23}) 
\be \label{8}
\hat{V}f=\ell_p^3\sum_{v\in V(\gamma)} \hat{V}_v f=\left( \ell_p^3\sum_{v\in 
V(\gamma)} \sqrt{|\frac{i}{8\cdot 3!}\sum_{e_I\cap e_J\cap e_K=v}
\epsilon(e_I,e_J,e_K)
\epsilon_{ijk}X^i_I X^j_J X^k_K|}\; \right) \;f_\gamma(g_1,..,g_n)
\ee
where $\epsilon(e_I,e_J,e_K)=
\mbox{sgn}(\det(\dot{e}_I(0),\dot{e}_J(0),\dot{e}_K(0)))$.
We have abbreviated $g_I=h_{e_I}(A)$ and $X_I=X(g_I)$ is the
right invariant vector field on $SU(2)$ (we have chosen orientations such
that all edges are outgoing at $v$). $V(\gamma)$ is the set of 
vertices of $\gamma$ and $\ell_p:=\sqrt{\hbar\kappa}$ is the Planck length. 
The symbol $\hat{V}_v$ is defined as follows :
$$
\hat{V}_v f_\gamma=\left\{ \begin{array}{lr}
\sqrt{|\frac{i}{8\cdot 3!}\sum_{e_I\cap e_J\cap e_K=v}
\epsilon(e_I,e_J,e_K)
\epsilon_{ijk}X^i_I X^j_J X^k_K|}\; \;f_\gamma
& \mbox{ if }v\in V(\gamma)\\
0 & \mbox{ if }v\not\in V(\gamma) 
\end{array}\right. ,
$$
that this, it 
is an operator which acts on the edges of any graph meeting at the point $v$
in the way displayed in (\ref{8}).
This demonstrates that $\hat{V}$
is a densely defined operator on thrice differentiable cylindrical 
functions. From
this it follows already that we have a chance of giving meaning to an 
operator corresponding to $H^E$.\\ 
Secondly, it is a well-known fact that $K$ is, up to a multiplicative 
constant, just the time derivative of 
the total volume with respect to the integrated Hamiltonian 
constraint (which is a signature invariant statement) 
\be \label{9}
K=-\{\frac{V}{\kappa},\int_\Sigma d^3x H^E(x)\} \ee 
which of course can also be verified immediately. So, if we (again) replace
$V,H^E$ by their quantizations and Poisson brackets by ($1/(i\hbar)$ times)
commutators we also find that we have a chance of giving meaning to an 
operator corresponding to $K$.\\
This completes our explanation of the main idea. The rest of this paper 
is devoted to a precise construction of the operators sketched in this 
section. We do this in a series of three steps :\\
Step A) We begin by giving meaning to an operator corresponding
to the Euclidean Hamiltonian constraint (\ref{7}). The result will be
a (self-adjoint) operator on $\cal H$ whose constraint algebra is 
anomaly-free.\\
Step B) We quantize $K$ along the lines sketched above using the
known quantizations $\hat{V},\hat{H}^E$.\\
Step C) We quantize the Wheeler-DeWitt constraint using the known
quantizations $\hat{H}^E,\hat{K}$ and exploiting (\ref{6}). We show that its
constraint algebra is anomaly free and that the result is a (symmetric)
densely defined operator on $\cal H$ as well.\\
Recalling that $C=(\pi/2)K$ is the classical generator of the Wick rotation
transform \cite{4} we naturally obtain its quantization, as well as that of 
the Euclidean Hamiltonian constraint, for free in our procedure.

\section{Quantization of the Euclidean Hamiltonian constraint}

The method applied in \cite{12,13} is to absorb the prefactor 
$1/\sqrt{\det(q)}$ in (\ref{7})
into the lapse function and to give meaning to the operator corresponding to
the square root of the trace. We will {\em not} do this. By employing the 
method described below we can avoid the complications that arise in 
connection with these two steps.
Also, it would be less straightforward to define $K$ with this 
rescaled form of 
the constraint since then $K$ is not the time derivative of the volume any 
longer.

\subsection{Regularization}

The regularization and all computations will be performed in an
arbitrary but fixed standard frame for $\Sigma$ as usual. The end result 
will be independent of that choice of frame.\\ 
We start from the classical expression for the Euclidean constraint 
functional 
\be \label{10}
H^E[N]=\frac{2}{\kappa}\int_\Sigma d^3x 
N(x)\epsilon^{abc}\mbox{tr}(F_{ab}\{A_c,V\})  \ee
where $N$ is the lapse function divided by $\kappa$.\\
We now triangulate $\Sigma$ into elementary tetrahedra $\Delta$ where 
each of its edges are analytic.
For each tetrahedron we single out one of its vertices and call it 
$v(\Delta)$. Let $s_i(\Delta),
\;i=1,2,3$ be the three edges of $\Delta$ meeting at $v(\Delta)$. Let 
$\alpha_{ij}(\Delta):=s_i(\Delta)\circ a_{ij}(\Delta)\circ 
s_j(\Delta)^{-1},\;\alpha_{ji}=\alpha_{ij}^{-1}$, be the loop based at 
$v(\Delta)$ where $a_{ij}$ is the obvious other edge of $\Delta$ connecting 
those endpoints of $s_i,s_j$ which are distinct from $v(\Delta)$.
Then it is easy to see that 
\be \label{10a}
H^E_\Delta[N]:=-\frac{2}{3}N_v\epsilon^{ijk}\mbox{tr}(h_{\alpha_{ij}(\Delta)}
h_{s_k(\Delta)}\{h_{s_k(\Delta)}^{-1},V\})
\ee
tends to the correct value 
$2\int_\Delta[N\mbox{tr}(F\wedge\{A,V\})]$ as we shrink $\Delta$
to the point $v(\Delta),\; N_v:=N(v(\Delta))$. Moreover, $H^E_\Delta[N]$ is 
manifestly gauge-invariant since $\alpha_{ij}\circ s_k\circ s_k^{-1}$ is 
a ``loop with a nose".\\ 
Let the triangulation be denoted by $T$. Then \be 
\label{11} H^E_T[N]=\sum_{\Delta\in T} H^E_\Delta[N]
\ee
is an expression which has the correct limit (\ref{10}) as all $\Delta$
shrink to their basepoints (of course the number of tetrahedra
filling any bounded subset of $\Sigma$ grows to infinity under this 
process).\\ As we have said before, we now simply replace $V$
by $\hat{V}$ and the Poisson bracket by $1/i\hbar$ times the commutator
and 
\be \label{12}
\hat{H}^E_T[N]:=\sum_{\Delta\in T} \hat{H}^E_\Delta[N],\;
\hat{H}^E_\Delta[N]:=-2\frac{N(v(\Delta))}{3i\ell_p^2}
\epsilon^{ijk}\mbox{tr}(h_{\alpha_{ij}(\Delta)}
h_{s_k(\Delta)}[h_{s_k(\Delta)}^{-1},\hat{V}])=:N_v \hat{H}^E_\Delta
\ee
is an operator with the correct classical limit.\\
It is obvious that the properties of the operator (\ref{12}) are largely
determined by the choice of triangulation $T$ and therefore we devote the
subsequent paragraph to a preliminary investigation of those properties
of (\ref{12}) which hold for any choice of triangulation. These
considerations will then  motivate our choices.

\subsubsection{Motivation}

The first property of (\ref{12}) that we wish to prove is that 
its action on cylindrical functions 
{\em is indeed finite} no matter how fine the triangulation $T$ is, provided
that a certain criterion is satisfied which we derive now.\\
In order to see this it is sufficient to consider the operator
$[h_{s_k(\Delta)},\hat{V}]f$ where $f$ is a cylindrical function with respect
to some graph $\gamma$. The first case is that 
$s_k(\Delta)\cap\gamma=\emptyset$. Then $V(\gamma\cup s_k(\Delta))=
V(\gamma)\cup V(s_k(\Delta))$ and it follows from (\ref{8})
that $[h_{s_k(\Delta)},\hat{V}]f=\sum_{v\in V(\gamma)} 
h_{s_k(\Delta)}\hat{V}_v f-\sum_{v\in V(\gamma\cup s_k(\Delta))} 
\hat{V}_v h_{s_k(\Delta)} f=-f\sum_{v\in V(s_k(\Delta))} 
\hat{V}_v h_{s_k(\Delta)}=0$ since $\hat{V}$ annihilates any cylindrical
function, whether gauge-invariant or not, whose underlying graph is not 
at least three-valent\footnote{We say that a graph is $n$-valent if there 
are no more than $n$ edges ingoing or outgoing at each vertex}.\\
The next case is that $s_k(\Delta)\cap\gamma\not=\emptyset$ but does not
contain a vertex of $\gamma$. That is, the set $V(\gamma\cup s_k(\Delta))
-V(\gamma)$
consists of vertices which are at most four-valent, however, the tangents 
of the edges incident at each of those vertices lie in a two dimensional 
vector space. It then follows that again $[h_{s_k(\Delta)},\hat{V}]f=
-\sum_{v\in V(\gamma\cup s_k(\Delta))-V(\gamma)}\hat{V}_v h_{s_k(\Delta)}f=0$
because of the signature factor $\epsilon(s_i,s_j,s_k)$ involved in 
(\ref{8}). Notice that this property would no longer be true had we used 
the volume operator as defined in \cite{22} : according to the second 
reference in \cite{23}, that operator does not vanish for vertices 
which involve only edges with co-planar tangents. Therefore, had we used 
this operator, as we make the triangulation finer and finer we would 
get more and more contributions and thus the resulting continuum
operator could not even be densely defined.\\ 
It follows that (\ref{12}) reduces to \be 
\label{12a} \hat{H}^E_T[N]f=\sum_{\Delta\cap V(\gamma)\not=\emptyset} 
\hat{H}^E_\Delta[N]f
=\sum_{v\in V(\gamma)}N_v\sum_{v \in \Delta} \hat{H}^E_\Delta f
=:\sum_{v\in V(\gamma)}N_v\hat{H}^E_v f
\ee
and we see that this expression is finite no matter how ``fine" the 
triangulation is, provided the number of tetrahedra intersecting the 
vertices of $\gamma$ stays bounded as we go to finer and finer 
triangulations. These considerations motivate to construct a triangulation
$T(\gamma)$ assigned to a graph $\gamma$ which meets this criterion and 
we get an operator $\hat{H}^E_{T(\gamma)}$. This furnishes our 
preliminary analysis.

\subsubsection{Requirements for a triangulation adapted to a graph}

Certainly, there are an infinite number of possible assignments. We choose
a particular assignment guided by the following principles :
\begin{itemize}
\item[a)] {\em Non-triviality} :\\ 
We could choose $T(\gamma)$ in such a way that there 
are no intersections with $\gamma$ at all, giving automatically a trivial 
result.
This is inappropriate as the number of degrees of freedom should be 
genuinely reduced by the Hamiltonian constraint.
\item[b)] {\em Diffeomorphism-Covariance} : \\
Remember that we want to impose the constraint on 
a distribution $\psi$ as outlined in section 2 and that the measure $\mu_0$ is
diffeomorphism invariant. This fact enables us to get rid 
of a huge amount of ambiguity arising in the assignment of a triangulation
as follows : the classical Hamiltonian constraint is not diffeomorphism 
invariant but it is diffeomorphism covariant. If we 
could carry over this classical property to the quantum theory then 
diffeomorphic vectors would be mapped by $\hat{H}^E$ into diffeomorphic 
vectors. Therefore, provided the state $\psi$ is diffeomorphism invariant,
the number $\psi[\hat{H}^E f]$ would only depend on the diffeomorphism
invariant properties of the triangulation assignment (this was first 
observed in \cite{12}).
The assignment should therefore move with the graph
$\gamma$ under diffeomorphisms of $\Sigma$.\\
More precisely, we have the following\footnote{This more precise 
formulation of the intuitive idea of diffeomorphism covariance was 
communicated to the author by Jurek Lewandowski} : Let $\gamma$ 
be a graph and $\phi(\gamma)$ its diffeomorphic image for some smooth
diffeomorphism $\phi\in\mbox{Diff}(\Sigma)$ (the diffeomorphism does not 
need to be analytic but it must keep the graph analytic). Then
$\hat{H}^E_{T(\gamma)}f_\gamma$ will depend on a graph $\widehat{\gamma}$
and likewise $\hat{H}^E_{T(\phi(\gamma))}f_{\phi(\gamma)}$ will depend on
a graph $\widehat{\phi(\gamma)}$. Then the requirement is that 
$\widehat{\gamma},\widehat{\phi(\gamma)}$ are diffeomorphic.\\
That is, we want that there exists $\phi'\in\mbox{Diff}(\Sigma)$ such that
$\hat{U}(\phi')[\hat{H}^E_{T(\gamma)}f_\gamma]=\hat{H}^E_{T(\phi(\gamma))}
f_{\phi(\gamma)}$ where $\hat{U}(\phi')f_\gamma=f_{\phi'(\gamma)}$ is 
a unitary representation of the diffeomorphism group 
$\mbox{Diff}(\Sigma)$ on $\cal H$. Notice that we have been dealing with
the Hamiltonian constraint at a point $\hat{H}^E$ rather than with its 
smeared version $\hat{H}^E[N]$ which is appropriate because we need to 
impose the constraint at every point of $\Sigma$. We will see that in 
our case the notion of a constraint operator at a point makes perfect 
sense.
\item[c)] {\em Cylindrical consistency} : \\
If $\gamma'$ is bigger than $\gamma$ then $T(\gamma)$
and $T(\gamma')$ will in general differ from each other. However, if $f$
is cylindrical with respect to $\gamma$ then the 
vectors $\hat{H}^E_{T(\gamma')}f$ and $\hat{H}^E_{T(\gamma)}f$ should be
diffeomorphic to each other. That is, we have cylindrical consistency
up to a diffeomorphism. The reason why we do not need to require exact 
cylindrical consistency is because the assignment of the triangulation is
only defined up to a diffeomorphism if we care only about the evaluation
of a diffeomorphism invariant state $\psi$ on the states 
$\hat{H}^Ef$\footnote{This observation again was first communicated to 
the author by Jurek Lewandowski}. 
\item[d)] {\em Symmetry and Self-Adjointness} :\\
The classical Hamiltonian constraint is a real-valued function on the 
phase space of general relativity. It is therefore compatible with the 
principles of quantum theory to construct an operator corresponding to it
which is self-adjoint or at least symmetric. While one is not forced to 
do so as the symmetric and non-symmetric operators have the same classical
limit and as we are only interested in the point $0$ of the spectrum, rather 
than the full spectrum, 
there are practical reasons, among others the applicability of 
the group averaging method and the possibility of being able to get 
rid of a quantization ambiguity, which motivate to have the 
constraint in 
symmetric form. We will therefore propose two quantizations of the 
Hamiltonian constraint : in the main text we will stick with a non-symmetric
operator as it turns out that it is technically much easier to handle and in
\cite{0} we will treat a symmetric version of the operator.
\item[e)] {\em Efficiency} : \\
The result of applying $\hat{H}^E_T[N]$ to cylindrical 
functions will be a cylindrical function that depends on additional edges.
We want to choose an assignment which introduces as less additional 
structure (edges) as possible.
\item[f)] {\em Naturality} : \\
The assignment should be uniform, that is, it treats all the
edges of the graph $\gamma$ incident at a vertex on equal footing.
\item[g)] {\em Anomaly-freeness} : \\
The assignment should be free of anomalies, that is,
the constraint algebra should close, otherwise we are reducing the number 
of degrees of freedom too much and we do not obtain a quantization of the 
classical theory.
\item[h)] {\em Non-Emptyness} :\\
The assignment should not be such that the kernel of the constraint 
operator is empty. The space of solutions to the classical Einstein equations
has a rich structure and so an empty kernel is not appropriate as it
would not correspond to the non-empty classical reduced phase space.
\end{itemize}

\subsubsection{Choice of a triangulation adapted to a graph}

The number of choices meeting these requirements is certainly still infinite.
The assignment $T(\gamma)$ we choose is as follows : we give in this 
subsection an assignment which is appropriate only for the non-symmetric 
regularization of the operator and will modify it at a later stage in
order to make it appropriate for the symmetric version.
\begin{itemize}
\item[0)] {\em Two-valent vertices} :\\ 
If a vertex $v$ is two-valent, 
adjoin one more edge to $\gamma$ incident
at $v$ and not intersecting $\gamma$ in any other point such that its tangent
at $v$ is transversal to all the tangents of edges of $\gamma$ at $v$. 
We will see later that the end result of the regularized 
operator is independent of that additional edge, even better, that 
functions on graphs with only divalent vertices are annihilated by 
$\hat{H}^E(N)$. \\
With this preparation we will assume from now on that all vertices 
are at least trivalent. Numerate all the edges 
$e_I$ of the (so possibly extended) graph by some index $I,J,K,...$.
Also we take all edges to be outgoing at a vertex as follows : by 
definition an analytical edge is an analytical embedding of a compact 
interval into $\Sigma$ and a vertex $v$ is a point of $\gamma$ such that 
there is no open neighbourhood $U\subset\Sigma$ of $v$ such that 
$\gamma\cap U$ is an embedded interval. By definition an edge is bounded 
by two vertices.  Given an edge $\tilde{e}$ of $\gamma$ with endpoints
$v,v'$ we subdivide it into two parts $\tilde{e}=e\circ e'$ where $e,e'$
respectively are outgoing at $v,v'$ respectively. Note that $e\cap e'$ 
is not a vertex of $\gamma$ so that $e,e'$ are strictly speaking no 
edges any longer but we will continue to call them edges again as it
simplifies the notation.
\item[1)] {\em Segments and arcs} :\\
Given an edge $e_I$ incident at a vertex $v$
choose $s_I$ to be a segment of $e_I$ which is such that it\\
a) is incident at $v$ with outgoing orientation and \\
b) which does not include the other endpoint $v_I$ of $e_I$ distinct from
$v$. \\
Also, given an unordered pair of edges $e_I,e_J$ incident at a vertex $v$, 
let $a_{IJ}$ be a curve which is such that that it \\
i) intersects $\gamma$ in the endpoints of $s_I,s_J$ distinct from $v$ and\\
ii) does not intersect $\gamma$ anywhere else.\\
The diffeomorphism invariant properties of the position of the arc 
$a_{IJ}$ will be specified more precisely below. We will see later 
that in order to meet the requirement that the constraint operator be 
symmetric on $\cal H$ we have to relax the prescriptions b),ii)
above.
\item[2)] {\em Tetrahedra saturating a vertex} :\\ 
For each vertex $v$ of $\gamma$ 
and each unordered triple of mutually distinct edges $(e_I,e_J,e_K)$
incident at $v$ 
(the number of such triples is given by $E(v)=n(v)(n(v)-1)(n(v)-2)/6$
where $n(v)$ is the valence of the vertex)
construct eight tetrahedra saturating $v$ as follows :\\
We have a map \\
$(s_I,s_J,s_K;a_{IJ},a_{JK},a_{KI})\to
(s_1(\Delta),s_2(\Delta),s_3(\Delta);
a_{12}(\Delta),a_{23}(\Delta),a_{31}(\Delta))$ 
where the labelling is such that the orientation of the the tangents at $v$ is
positive and we have indexed these six segments by the obvious 
tetrahedron $\Delta$ that they form. Choose the basepoint of $\Delta$ to be 
the vertex of $\gamma$ under consideration, that is, $v(\Delta)=v$.\\
We are now ready to construct the eight tetrahedra saturating $v$ for the
triple $(e_I,e_J,e_K)$ :\\
Let $[0,1]\to s_i(\Delta)(t)$ and $[0,1]\to a_{ij}(\Delta)(t)$ be 
parameterizations of $s_i(\Delta)$ and $a_{ij}(\Delta)$ respectively. Define 
their ``mirror images" by\\
$s_{\bar{i}}(\Delta)^a(t):=2v^a-s_i(\Delta)^a(t),$\\
$a_{\bar{i}\bar{j}}(\Delta)^a(t):=2v^a-a_{ij}(\Delta)^a(t),$\\
$a_{\bar{i}j}(\Delta)(t):=a_{\bar{i}\bar{j}}(\Delta)^a(t)
-2t[v-s_j(\Delta)(1)]^a,$\\ 
$a_{i\bar{j}}(\Delta)(t):=a_{ij}(\Delta)^a(t)
+2t[v-s_j(\Delta)(1)]^a$\\
respectively
where $v^a$ are the coordinates of the vertex $v$. Here we have assumed 
that all objects lie in a chart containing $v$ which is always possible 
by choosing the basic quantities $s_i(\Delta),a_{ij}(\Delta)$ small 
enough. Notice that by definition $s_i(\Delta)(0)=v,
s_i(\Delta)(1)=a_{ij}(\Delta)(0),s_j(\Delta)(1)=a_{ij}(\Delta)(1)$ so that
$s_{\bar{i}}(\Delta)(0)=v,
s_{\bar{i}}(\Delta)(1)=a_{\bar{i}\bar{j}}(\Delta)(0)=a_{\bar{i}j}(\Delta)(0),
s_{\bar{j}}(\Delta)(1)=a_{\bar{i}\bar{j}}(\Delta)(1)=a_{i\bar{j}}(\Delta)(1),
s_i(\Delta)(1)=a_{i\bar{j}}(\Delta)(0),s_j(\Delta)(1)=a_{\bar{i}j}(\Delta)(1)$.
We did not make use of any background metric. We now form loops
$\alpha_{\bar{i}\bar{j}},\alpha_{\bar{i}j},\alpha_{i\bar{j}}$ and combine 
them with $s_k,s_{\bar{k}}$ to form seven more right oriented tetrahedra. 
Together with $\Delta$ these are the eight tetrahedra that 
we looked for. We will see that the seven ``mirror" tetrahedra do not play 
any role at the end of the day so that the choice of adapted frame is 
irrelevant to define them.

Although we will only be concerned later with the one-dimensional edges of 
the tetrahedra constructed, we will need also their two- and 
three-dimensional properties in an intermediate step : \\
Choose any surfaces bounded by the edges of these eight tetrahedra and 
define 
the closed subset of $\Sigma$ bounded by those faces of a tetrahedron 
$\Delta$ to be the region assigned to $\Delta$.
The only property of the region assigned to
$\Delta$ which will be important is that $\Delta$ and its mirror images 
saturate $v$ which is a diffeomorphism invariant property and therefore
everything will be independent of these regions.
\item[3)] {\em Diffeomorphism invariant prescription for the position of the 
arcs $a_{ij}(\Delta)$} :\\ 
The following lemma, whose proof to the best of our knowledge is 
unpublished, shows that one can always choose two curves to lie in 
the $x/y$ plane of an adapted coordinate system in which they take a 
standard form.
\begin{Lemma}
Let $s_1,s_2$ be two distinct analytic curves which intersect only in their
starting point $v$. There exist parameterizations of these curves, a 
number $\delta>0$ and an analytic diffeomorphism 
such that in the corresponding frame the curves are given by\\
a) $s_1(t)=(t,0,0),\; s_2(t)=(0,t,0),\;t\in [0,\delta]$
if their tangents are linearly independent at $v$ and\\
b) $s_1(t)=(t,0,0),\; s_2(t)=(t,t^n,0),\;t\in [0,\delta]$
for some $n\ge 2$ if their tangents are co-linear at $v$.\\
We will call the associated frame a frame adapted to $s_1,s_2$.
\end{Lemma}
Proof :\\
Given a frame, denote by $b_i,i=1,2,3$ the standard vector with entry $1$
at the i-th index and zero otherwise.\\
First we show that any curve $s$ can be mapped into a straight line by an
analytic diffeomorphism. To that end, let us expand 
$s(t)=f^i(t)b_i$ where $f^i$ are analytic functions of $t$. Since
$\dot{s}$ is nowhere vanishing, at least one of the functions, say $f^1$,
has the property $\dot{f}^1(0)\not=0$ and so it does not in a neighbourhood
of $0$. Choose $b_1':=b_1,b_1-b_2,b_1-b_3,b_1-b_2-b_3$ and 
$(f^{2\prime},f^{3\prime})=(f^2,f^3),(f^2+f^1,f^3),(f^2,f^3+f^1),
(f^2+f^1,f^3+f^1)$ whenever $(\dot{f}^2(0),\dot{f}^3(0))$ are
$(\not=0,\not=0),(=0,\not=0),(\not=0,=0),(=0,=0)$. We conclude that
we can write $s(t)=f^1(t)b_1'+(f^2)'(t)b_2+(f^3)'(t)b_3=:g^i(t) b_i'$ where 
$b_i'$ form a basis 
and $\dot{g}_i(0)\not=0$. It follows by the inverse function theorem 
that the equation $x^i=g^i(t)$ can be inverted in a neighbourhood of
$0$ and that $(g^i)^{-1}(x^i)$ is analytic in a neighbourhood of $0$
because $g^i(t)$ is of order $o(t)$. We now see that the following 
diffeomorphism $x^{i\prime}(x^1,x^2,x^3):=(g^i)^{-1}(x^i)$ is analytic
and maps $s(t)$ to $s'(t)=x'(s(t))=t(b_1'+b_2'+b_3')$. Upon performing
a constant diffeomorphism (that is, a $GL(3)$ transformation) we can 
achieve that $s'(t)=t b_1$.\\
So we can assume that we have already mapped $s_1,s_2$ so that
$s_1(t)=t b_1$. Now consider $s_2(t)=f^i(t)b_i$.\\
Case a) Since $\dot{s}_1(0),\dot{s}_2(0)$ are not co-linear (which is a
diffeomorphism invariant statement) it follows that not both of
$\dot{f}^2(0),\dot{f}^3(0)$ can vanish. So let us assume that for instance
$\dot{f}^2(0)\not=0$. By a similar argument as above we can write
$s_2(t)=g^1(t)b_1+g^2(t) b_2'+g^3(t) b_3$ where $b_1,b_2',b_3$ are linearly
independent and $\dot{g}^i(0)\not=0$. The analytic diffeomorphism
$x^{i\prime}(x^1,x^2,x^3)=(g^i)^{-1}(x^i)$ maps
$s_1'(t)=x'(s(t))=g^{1\prime}(t)b_1,\;s_2'(t)=t(b_1+b_2'+b_3)$.
Now, a change of parameterization $t'=g^{1\prime}(t)$ for $s_1$ and
a final $GL(3)$ transformation proves the assertion.\\
Case b) Since $\dot{s}_1(0),\dot{s}_2(0)$ are co-linear it follows that
$\dot{f}^1(0)\not=0,\dot{f}^2(0)=\dot{f}^3(0)=0$. Let $n\ge 2$ be the smaller
of the two numbers $n_2,n_3$ defined by $f^2(t)=o(t^{n_2}),f^3(t)=o(t^{n_3})$
($n$ is finite because the curves are not identical).
Without loss of generality we may assume $n_2=n$ and now it follows
by an already familiar argument that we can write 
$s_2(t)=g^1(t)b_1+g^2(t)b_2'+g^3(t)b_3$ where $b_1,b_2',b_3$ are 
linearly independent and $\dot{g}^1(0),(g^2)^{(n)}(0),(g^3)^{(n)}(0)\not=0$.
It follows that $g^2(t)=:\tilde{g}^2(s),g^3(t)=:\tilde{g}^3(s)$ are analytic 
functions of the analytic coordinate $s:=t^n$ and that they are 
invertible in a neighbourhood of $0$. We now define the analytic
diffeomorphism $x^{i\prime}(x^1,x^2,x^3):=(\tilde{g}^i)^{-1}(x^i)$ where
$g^1=\tilde{g}^1$ which maps 
$s_1'(t)=b_1 (g^1)^{-1}(t),\;s_2'(t)=b_1 t+t^n(b_2'+b_3)$ and again 
a reparameterization for $s_1$ and a $GL(3)$ transformation proves the 
assertion.\\
$\Box$\\
Using this lemma
we will now describe more precisely the choice of the arcs $a_{ij}(\Delta)$.
We follow and extend (to the case of a pair of edges with co-linear
tangents at their intersection) an elegant
prescription due to Jurek Lewandowski \cite{13} to which one is driven quite
naturally. For the sake of self-containedness of the present paper we repeat 
the argument here to the extent we need it. The discussion is rather
technical and lengthy and the reader not interested in the details may 
skip the rest of the present item and just should assume that there 
exists a diffeomorphism invariant prescription of the topology of the 
routing.\\

Let $s_1,s_2$ be two segments of edges $e_1,e_2$ of $\gamma$, incident at 
the vertex $v$. Their other endpoints $v_1,v_2$ are connected by an arc $a$. 
Basically we wish to avoid that the arc $a$ intersects any other point of 
$\gamma$ different from $v_1,v_2$.
Clearly, by choosing $a$ to lie in a small enough neighbourhood of $v$
the only danger is that $a$ can intersect some other edge $e$ different 
from $e_1,e_2$, also incident at $v$. With rising valence of $v$ the 
number of topologically different possibilities of routing $a$ between the 
edges $e$ incident at $v$ becomes rather complex so that we need to give 
a diffeomorphism invariant description of the choice of routing.\\
By choosing a frame adapted to $s_1,s_2$ we have partially fixed the  
gauge freedom associated with analyticity preserving diffeomorphisms. 
Consider any other frame $(x^{i\prime})$ connected to the identity 
(so that in particular $x^{i\prime}_{,x^i}>0$)
and preserving the adaptedness of the frame. 
Thus $x^{i\prime}(s_1(t))=(t,0,0)$ implies that, upon taking arbitrary
derivatives with respect to $t$, that $\partial^m_x y'=\partial^m_x z'=0$
for all $m>0$ at $v$. Likewise, $x^{i\prime}(s_1(t))=(0,t,0)$ implies 
$\partial^m_y x'=\partial^m_y z'=0$ for all $m>0$ at $v$. However, for
$x^{i\prime}(t,t^n,0)=(t,t^n,0)$ it is more subtle to characterize the 
partial derivatives. Consider the $k-$th derivative at $t=0$ of 
the function $z'(t,t^n,0)=0$. Let $k=mn+r$ where $r=0,..,n-1$. It follows 
that we get a linear combination, with positive coefficients, of the 
following partial derivatives at $v$ : $\partial^{r+ln}_x\partial^{m-l}_y
z'$ for $l=0,..,m$ which has to vanish. In particular, the case 
$m=1,r=0,..,n-1$
shows that we get a linear combination of $\partial^{r+n}_x z'$ and 
$\partial^r_x\partial_y z'$ so that $\partial^r_x\partial_y z'=0$ at $v$
for all $r=0,..,n-1$. The case $m=2,r=0$ gives a linear combination 
of $\partial^{2n}_x z'=0,\partial^n_x\partial_y z',\partial^2_y z'$ and 
the latter two terms do not need to vanish separately. Working out the 
precise coefficients of that linear combination we find that 
$$
n! \left( \begin{array}{c} 2n \\ n 
\end{array} \right)\partial^n_x\partial_y z' +(n!)^2 
\left( \begin{array}{c} 2n-1 \\ n \end{array} 
\right) \partial^2_y z'=0\;.
$$
It turns out that this is all we need to prescribe the routing of the arc.\\
The idea is that we wish to choose the arc $a$ to lie in the coordinate plane
bounded by $s_1,s_2$ (that is, either in the quadrant $x,y\ge 0$ or 
the wedge
$\{(x,y,z);\;0\le y\le x^n,\;z=0\}$ respectively) and the question is under 
which 
circumstances we can guarantee that then $a$ does not intersect $e$. 

The easiest case
is that $\dot{e}^z(0)\not=0$. We show that $\mbox{sgn}(\dot{e}^z(0))$ is
preserved under change of adapted frames. Namely
$\dot{e}^{z'}(0)=z'_{,a}\dot{e}^a(0)=z'_{,z}\dot{e}^z(0)$. By analyticity
it then follows that $e$ is curved away from the $x/y$ plane in a 
neighbourhood of $v$ and so does not intersect $a$ for no choice of 
adapted frame and therefore for no frame.

The case $\dot{e}^z(0)=0$ is more tricky. Without loss of generality we may
assume that at least $\dot{e}^x(0)\not=0$ (otherwise $\dot{e}(0)=0$;
switch the role of $x$ and $y$ if necessary for the case that $s_1,s_2$ 
do not have co-linear tangents at $v$. If they do have co-linear 
tangents at $v$ and $\dot{e}^x(0)=0$ then $\dot{e}^y(0)\not=0$ and it 
follows that a segment incident at $v$ of the projection of the curve $e$ 
into the $x/y$ plane lies above the parabola $y=x^n$ which describes $s_2$
and so $a$ cannot intersect $e$ anyway). 

We can make a distinction between two 
situations.\\ Situation A : There exists a finite number $m'$ such that 
$(e^{(m')})^z(0)\not=0$
and $m\ge m',\;m\le\infty$ such that $(e^{(m)})^y(0)\not=0$ are the first 
non-vanishing derivatives. The combination $m'=m=\infty$ is excluded as 
otherwise $s_1$ and $e$ would overlap in a finite segment (here we have 
used the analyticity of the edges). We then readily
verify that under a change of adapted frame we have that the first 
non-vanishing derivative of the $z$-component of $e$ is given by
$(e^{(m')})^{z'}(0)=z'_{,z}(e^{(m')})^z(0)$ so that the sign of 
$(e^{(m')})^z(0)$
is again preserved and $e$ is curved away from the $x/y$ plane so as not 
to intersect $a$.\\
Situation B : There exists a finite number $m$ such that 
$(e^{(m)})^y(0)\not=0$
and $m'>m,\;m'\le\infty$ such that $(e^{(m')})^z\not=0$ are the first 
non-vanishing derivatives. Again $m=m'=\infty$ is an excluded possibility.

{\em Case that $\dot{s}_1(0),\dot{s}_2(0)$ are not co-linear.}\\
If $e$ does not point into an octant of the frame where both $x/y$ are 
positive then again there is no danger that $a$ intersects $e$ since we 
choose $a$ to lie in the $x/y$ plane with positive $x/y$ components as said
above. 
Since $x'_{,x},y'_{,y}\not=0,\partial^k_x y'=\partial^k_y x'=0\forall k$ it 
follows that the sign of the first non-vanishing derivatives of $e^x,e^y$ 
are preserved under changes of the adapted frame. \\
So let us assume that $e$ does point into an octant where both $x/y$
are positive. This means that $\dot{e}^x(0),\;(\dot{e}^{(m)})^y(0)>0$.

The first possibly non-vanishing derivative of the $z$-component of $e$
is given by
$(e^{(m+1)})^{z'}(0)=z'_{xy}\dot{e}^x(0)(\dot{e}^{(m)})^y(0)+
z'_{,z}(e^{(m+1)})^z(0)$
and we see that this can have any sign for a suitable choice of $z'$.
We use this freedom to further fix the frame such that this sign is positive.
Choose $z'(z,y,z)=z+\beta xy$. This satisfies all requirements on $z'$
at $x=y=z=0$ and we see that upon choosing one and the same $\beta(s_1,s_2)$ 
large 
enough we can beat the terms $z'_{,z}(e^{(m+1)})^z(0)$ for an arbitrary (but 
finite) number of edges $e$ as to make $(e^{(n+1)})^{z'}(0)>0$.

{\em Case that $\dot{s}_1(0),\dot{s}_2(0)$ are colinear.}\\
If $e$ does not point into the wedge $y\le x^n,x\ge 0$ then again the arc
$a$ cannot possibly intersect $e$ if we choose it close enough to the vertex.
This time we only have that $x'_{,x},y'_{,y}\not=0,\partial^k_x y'\forall 
k$ which implies only that the sign of the first non-vanishing derivative 
of $e^y$ is preserved, we even find that $x'_{,x}=y'{,y}=1$ so that even 
its value is preserved. Now, if $m\ge 2$ then the sign of $\dot{e}^x(0)$
is preserved. If $m=1$ then it is not necessarily preserved but then 
it is true that, since $n\ge 2$, the projection of $e$ into the $x/y$
plane, which lay outside the wedge in a small enough neighbourhood of the 
vertex, is still outside the wedge. Therefore the condition that $e$
does or does not point into the wedge is preserved. So let us assume that
$e$ points into the wedge, in particular,
$\dot{e}^x(0),\;(\dot{e}^{(m)})^y(0)>0$.\\
We notice that the first derivative of $e^{z'}$ at $t=0$ which involves 
$A:=\partial_x^n\partial_y z'$ at $v$ is of order $n+m$ and given
by a term $c_A A (\dot{e}^x(0))^n (e^{(m)})^y(0),\;c_A>0$. Likewise,
the first derivative which involves $B:=\partial_y^2 z'$ at $v$ is of order
$2m$ and given by a term $c_B B [(e^{(m)})^y(0)]^2,\;c_B>0$.\\
Subcase I) $m'\le\min(2m,m+n)$\\
In this case the first non-vanishing derivative of $e^{z'}$ at $t=0$ is 
of order $m'$ and involves a term $z'_{,z}(e^{(m')})^z(0)$. By choosing
the coefficient $z'_{,z}$ large and positive enough we can beat any possible 
contribution involving $A,B$ and preserve the sign of $(e^{(m')})^z(0)$.\\
Subcase II) $m'>\min(2m,m+n)$\\
The first non-vanishing derivative of $e^{z'}$ at $t=0$ is 
of order $m+n$ if $n<m$ and involves only the term proportional to
$A$, of order $2m$ if $n\ge m$ and involves only a term proportional to
$B$ if $n>m$ and for $n=m$ the derivative is precisely given by
$$
(e^{(2n)})^{z'}(0)=B (e^{(n)})^y(0) \left( \begin{array}{c} 2n-1\\n 
\end{array} \right) [(e^{(n)})^y(0)-n!(\dot{e}^x(0))^n]
$$
where we have made use of the relation between $A,B$ as stated above.
By choosing $e^x(t)$ as a parameter we may assume that $e^x(t)=t$. Then
we may assume $(e^{(n)})^y(0)<n!$ : Namely, if $(e^{(n)})^y(0)>n!$ then
again a segment incident at $v$ of the projection of the curve $e$ into 
the $x/y$ plane, whose
$y-$component looks like $e^y(t)=t^n+o(t^{n+1})$, lies above the parabola
$y=x^n$ which describes $s_2$ and cannot possibly intersect the arc $a$.
If $(e^{(n)})^y(0)=n!$ then we must have that either $e^y$ has 
another higher order non-vanishing derivative $(e^{(m)})^y(0)\not=0,m>n$
or, if that is not the case, $m'<\infty$ for otherwise $e$ would coincide 
with $s_2$ which we excluded. In the latter case, the first non-vanishing
derivative of $e^{z'}$ at $t=0$ would be again $z'_{,z}(e^{(m')})^z(0)$ 
because by definition of a diffeomorphism preserving the adaptedness of 
the frame, all derivatives of $z'(e(t))=z'(t,t^n,e^z(t))$ at $t=0$ which 
do not involve at least one partial derivative with respect to $z$ must
vanish. So again the sign of $(e^{(m')})^z(0)$ would be preserved. In the
former case, the first non-vanishing derivative of 
$e^{z'}$ at $t=0$ and which is proportional to $\partial^k_x\partial^l_y$
for some $k\ge0$ and some $l>0$ at $v$ is of order $m+n$ and proportional to
$\partial^n_x\partial_y z' (e^{(m)})^y(0)$ which follows from the fact 
that all contributions which do not contain at least one factor of 
$(e^{(m)})^y(0)$ must vanish due to the adaptedness of the frame. Now we 
are back to either $m'\le n+m=\min(m+n,2m)$ or $m'>m+n$ and we have the case 
$m>n$.\\
It follows then that if we choose any positive number $\beta$ and some
large enough positive number $\alpha>0$ and the diffeomorphism (which
satisfies the condition between $A,B$ as given above)
$z':=\alpha z+\beta(y^2/2-2y x^n/(n!)^2)$ then the image of the edge
$e$ under this diffeomorphism will be such that it is curved away from the
$x/y$ plane as before for $m'\le \min(m+n,2m)$ (where it is understood that
we take $m$ to be the next to leading order of $e^y(t)$, with positive 
coeeficient, in case 
that $e^x(t),e^y(t)=t^n+o(t^m)$ and $m=\infty$ otherwise) and for $m'>
\min(m+n,2m)$ it is curved into the upper half space if $n>m$, into the 
lower half space if $n<m$ and into the lower half space if 
$e^x(t)=t,e^y(t)=kt^n+o(t^{n+1}),\;k<1$ (for the other cases 
there is no routing to be chosen). This can be achieved for an 
arbitrary number of edges $e$ with the same 
$\alpha(s_1,s_2),\beta(s_1,s_2)$. Notice that the topology of the 
routing is diffeomorphism invariantly defined as the numbers $n,m,m'$
are diffeomorphism invariant.\\  

Concluding, we choose an adapted frame and a small enough 
neighbourhood of $v$ and an arc $a_{ij}(\Delta)$ going through that 
neighbourhood, which lies in the plane bounded by $s_i(\Delta),s_j(\Delta)$
such that the routing through all other edges incident at $v$ is the one
described above.\\
\item[4)] {\em Tetrahedra away from the vertices} :\\
Denote by $D(\Delta)$ the closed region in $\Sigma$ filled by the eight
tetrahedra constructed from a triple $e_I,e_J,e_K$ as outlined in 1).
Also consider their union 
$D(v)=\cup_{v(\Delta)=v} D(\Delta)$ and their complement with respect to 
$D(v)$, that is, $\bar{D}(\Delta):=D(v)-D(\Delta)$. We triangulate
$D(\Delta)$ as outlined in 1) by the eight tetrahedra constructed.  
The sets $\bar{D}:=\Sigma-\cup_{v\in V(\gamma)} D(v)$ 
and $\bar{D}(\Delta)$ are triangulated arbitrarily.
As we have argued above, the final result will not depend on 
these tetrahedra because they do not intersect a vertex of $\gamma$. 
This follows from the fact that all the tetrahedra different from 
the ones constructed in 1) have a basepoint different from any of the 
vertices of $\gamma$ since by construction the tetrahedra described in 1)
saturate them.
\item[5)] {\em Closeness to a triangle} :\\
Having fixed the topology of the routing of the arcs as in 2) 
we now choose, in the standard frame, the arc $a_{ij}(\Delta)$ to be 
as straight as possible so that $\alpha_{ij}(\Delta)$ looks like a 
triangle,
as much as the routing allows. This will then justify the 
approximation $h_{\alpha_{ij}(\Delta)}\approx 1+\frac{(\delta t)^2}{2}
\dot{s}_i(\Delta)^a(0)\dot{s}_j(\Delta)^b(0)F_{ab}(v(\Delta))$ where
$[0,\delta t]$ is the subinterval of $[0,1]$ corresponding to 
$s_i(\Delta)$ as compared to the whole edge of $\gamma$ of which it is a 
segment. 
\end{itemize}

\subsection{Final Regularization of the Euclidean constraint}

Let us now write the integral over $\Sigma$ (\ref{10}) for the classical 
theory as follows 
\ba \label{13}
&&\int_\Sigma=\int_{\Sigma-\cup_{v\in V(\gamma)} D(v)}+\sum_{v\in V(\gamma)} 
\int_{D(v)} =\int_{\bar{D}}+\sum_{v\in V(\gamma)} \frac{1}{E(v)}
\sum_{v(\Delta)=v}[\int_{D(\Delta)}+\int_{\bar{D}(\Delta)}]
\nonumber\\
&& =\sum_{\Delta'\in\bar{D}}\int_{\Delta'}+\sum_{v\in V(\gamma)} 
\frac{1}{E(v)}\sum_{v(\Delta)=v}
[\sum_{\Delta'\in\bar{D}(\Delta)}\int_{\Delta'}+
\sum_{\Delta'\in D(\Delta)}\int_{\Delta'}] \;. \ea
In the classical expression (\ref{11}) we can replace, for each $\Delta$,
$\sum_{\Delta'\in D(\Delta)}\int_{\Delta'}$ by $8\int_\Delta$
as all the tetrahedra shrink to their basepoints so that the dependence
on the mirror images of $\Delta$ drops out as promised.\\
Now we quantize the resulting expression (\ref{11}) as outlined above and
it follows from the considerations 
in section 3.1.1 
that on the cylindrical 
subspace labelled by the graph $\gamma$ the action of the Euclidean 
Hamiltonian constraint reduces to
\be \label{14}
\hat{H}^E_{T(\gamma)}[N]f:=\hat{H}^E_\gamma[N] f
=\sum_{v\in V(\gamma)}N_v \frac{8}{E(v)}\sum_{v(\Delta)=v} 
\hat{H}^E_\Delta f =:\sum_{v\in V(\gamma)}N_v\hat{H}^E_v f
\ee
This furnishes the regularization step. In particular, the expression 
(\ref{14}) is finite (i.e. a cylindrical function) independently of 
$T$ because the number $E(v)$ is determined by the graph 
and not by the triangulation and so does not change as we make the 
triangulation finer and finer.\\
Let us now show that divalent vertices do not contribute :
if $v$ is a divalent vertex of $\gamma$, if $s_3$ is the segment 
of the edge added in requirement 0) and if $\alpha_{12}$
is the corresponding loop along segments $s_1,s_2$ of the two edges 
$e_1,e_2$ of $\gamma$ incident at $v$ then
\be \label{14a}
\hat{H}^E_v=-\frac{16}{3i\ell_p^2}
\mbox{tr}((h_{\alpha_{12}}-h_{\alpha_{12}}^{-1})
h_{s_3}[h_{s_3}^{-1},\hat{V}]) \;. 
\ee
The terms involving $\alpha_{23},\alpha_{31}$ drop out
which follows from the fact that the volume operator annihilates divalent
vertices as is obvious from the expression (\ref{8}). Notice that if 
$s_3$ was not transversal to $s_1,s_2$ at $v$ then (\ref{14a}) would vanish
trivially because the volume operator annihilates vertices which are such 
that all incident edges have co-planar tangents. But even so, (\ref{14a}) 
vanishes : 
since the end result of applying (\ref{14a}) to a gauge invariant 
function $f$ cylindrical with respect to a graph $\gamma$
must be gauge invariant, it cannot depend on $s_3$ (to see this, expand 
$\hat{H}^E_v f$ into spin-network states. Each of these spin-network 
states can colour $s_3$ only with spin $0$ because there is no edge 
of $\gamma$ available in order to combine with the other endpoint of
$s_3$ different from $v$ in a gauge invariant way). Moreover, it is 
easy to see that only a term proportional to 
$h_{s_3}\hat{V} h_{s_3}^{-1}$
survives. Now from 
the fact that the space of vertex contractors for divalent vertices is
one-dimensional we see that $h_{s_3}\hat{V} h_{s_3}^{-1}f=\lambda f$
so that $\mbox{tr}([h_{\alpha_{12}}-h_{\alpha_{12}}]h_{s_3}\hat{V} 
h_{s_3}^{-1})f=\lambda\mbox{tr}([h_{\alpha_{12}}-h_{\alpha_{12}}])f
=0$ where we used the $SU(2)$ Mandelstam identity. The argument 
actually extends to the case of vertices of arbitrary valence but such 
that all incident edges have co-planar tangents. So we see 
already that functions cylindrical on graphs with only such degenerate 
vertices are annihilated by $\hat{H}^E(N)$. This is a feature which
is shared with previous regularizations \cite{26,27}. \\

It is amazing that one got expression (\ref{14}) almost for free once
one knows that the volume operator is well-defined on holonomies,
no ill-defined products of distributions arise, we do not encounter
any singularities, no renormalization of the operator is necessary.
Note that we have no problems 
in ordering $\hat{V}$ to the left or to the right of the holonomies 
involved as $\hat{V}$ has a finite action on holonomies of $A$ as is 
clear from (\ref{8}).

\subsection{Cylindrical consistency}

We have now produced an uncountable family of Euclidean Hamiltonian 
constraint operators $(\hat{H}^E_\gamma(N))_\gamma$, one operator for 
each graph $\gamma$ given in (\ref{14}). What we need to make sure is that 
these operators 
are the projections to cylindrical subspaces of one and the same 
operator on $\cal H$. This requirement will lead to some 
modifications of (\ref{14}) (while keeping the classical limit to be 
still $H^E(N)$).\\ 
Also we would like to construct one version of $\hat{H}^E(N)$  
which is symmetric. The way it stands, not even the projections 
of $\hat{H}^E(N)$ given in (\ref{14}) are symmetric operators on $\cal H$. 
Therefore, for the symmetric operator, we first order each term associated 
with a tetrahedron symmetrically . The 
result is (using that $\hat{V}$ is symmetric on $\cal 
H$ and $h_e^\dagger=\overline{h_e}=(h_e^T)^{-1}$ on $\cal H$) that, in 
case we wish to construct a symmetric operator, we replace (\ref{14}) by 
\ba \label{15}
&& \hat{h}^E_\gamma[N] f:=\sum_{v\in\gamma} 
N_v \hat{h}^E_v 
f,\;\hat{h}^E_v=\frac{8}{E(v)}\sum_{v(\Delta)=v} 
\hat{h}^E_\Delta,\;
\nonumber\\
&& \hat{h}^E_\Delta:=-\frac{1}{3i\ell_p^2}
\epsilon^{ijk}\mbox{tr}(\{h_{\alpha_{ij}(\Delta)},
h_{s_k(\Delta)}[h_{s_k(\Delta)}^{-1},\hat{V}]\})
\ea
where $\{.,.\}$ denotes the anti-commutator, while we stick with
(\ref{14}) if we do not wish to construct a symmetric operator. With this 
choice, 
each $\hat{h}^E_\gamma[N]$ in (\ref{15}) separately becomes a symmetric 
operator on $\cal H$. Notice that this is far from guaranteeing that 
$\hat{H}^E(N)$ itself is symmetric (see \cite{0}). 
Both operators in (\ref{14}), (\ref{15}) have the 
dense domain $\mbox{Cyl}^3(\agb)$ inherited from 
$\hat{V}$ \cite{23}, the thrice differentiable cylindrical functions on 
$\agb$. \\
We now come to make both (\ref{14}),(\ref{15}) cylindrically consistent.
Let $\Delta_e=X^i(h_e)X^i(h_e)$ be the Casimir operator associated with 
an edge $e$ of $\gamma$ incident at $v$. Note that since $X(h_e)$ is 
right-invariant we have
$X(h_e)=X(h_s)$ for any segment $s$ of $e$ as long as $s$ is incident 
at the outgoing endpoint of $e$. We can now define an edge projector
$\hat{p}_e:=\theta(\hat{j}_e)$ where $\hat{j}_e:=\sqrt{1/4-\Delta_e}-1/2$
has spectrum $0,1/2,1,3/2,..$ and $\theta$ is some smooth function on
$\Rl$ which vanishes on $(-\infty,1/8])$ and equals $1$ on $[3/8,\infty)$.
The operators $\hat{p}_e$ are all commuting among each other and 
symmetric. The effect of this operator when applied to a function $f$
cylindrical with respect to a graph $\gamma$ is to annihilate that function
if $\gamma$ and $e$ do not intersect in a finite segment incident at the 
outgoing endpoint of $e$ and to leave it invariant otherwise. 
From these projectors we construct a tetrahedron projector 
$\hat{p}_\Delta:=\hat{p}_{s_1(\Delta)}\hat{p}_{s_2(\Delta)}
\hat{p}_{s_3(\Delta)}$ and a vertex operator 
$\hat{E}(v):=\sum_{v(\Delta)=v}\hat{p}_\Delta$.
We then define the following self-consistent (up to a 
diffeomorphism) family of (symmetric) operators 
\be \label{16}
\hat{H}^E_\gamma[N]:=\sum_{v\in V(\gamma)}8 N_v\sum_{v(\Delta)=v}
\left\{ \begin{array}{l}
\hat{H}^E_\Delta
\frac{\hat{p}_\Delta}{\hat{E}(v)}\mbox{ for (\ref{14})}\\
\frac{\hat{p}_\Delta}{\sqrt{\hat{E}(v)}}\hat{h}^E_\Delta
\frac{\hat{p}_\Delta}{\sqrt{\hat{E}(v)}}\mbox{ for (\ref{15}).}
\end{array} \right.
\ee
We note that (\ref{16}) still has the correct classical limit : up to 
terms of higher order in $\hbar$ its action on cylindrical functions is 
still given by (\ref{14}) which was shown to have the correct classical 
limit.\\
The symmetry of each member of the family defined in the second line of 
(\ref{16}) is obvious. The self-consistency 
of both families defined in (\ref{16}) can be checked as follows :\\
A graph $\gamma\subset\gamma'$ can be obtained from $\gamma'$ by a 
finite sequence of steps consisting of the following two basic ones : a) 
delete
an edge of $\gamma'$ and b) after removing an edge $e$, if one (or both)
of the former endpoints $v$ of $e$ is now divalent and $v=e_1\cap e_2$
where $e_1,e_2$ are analytic extensions of each other then combine
$e_{12}:=e_1\circ e_2^{-1}$ to an edge of $\gamma$, that is, delete a
vertex. In case a) it is clear that each term in
(\ref{16}) corresponding to a tetrahedron which involves the removed edge 
vanishes identically while in case b) all terms corresponding to the 
removed vertex vanish when applied to a function $f$ cylindrical with 
respect to the graph $\gamma$. Moreover, $\hat{E}(v)$ reduces to the correct 
value on $f$. Finally, it follows from our manifestly
diffeomorphism-invariant prescription of a loop-assignment that 
$\hat{H}^E_{\gamma'}f$
and $\hat{H}^E_\gamma f$ are related by a diffeomorphism. In more detail,
we have the following :
if $\gamma'$ is bigger than $\gamma$ and if $e_1',e_2'$ are edges of 
$\gamma'$ which are the parts of
the edges $e_1,e_2$ of $\gamma$ incident at the same vertex $v$ at which 
$e_1,e_2$ are incident then the loop made from the corresponding segments
$s_1',s_2'$ and $s_1,s_2$ are diffeomorphic thus guaranteeing 
cylindrical consistency up to a diffeomorphism.
This is enough to show consistency.\\
In the sequel we will prove that both operators in (\ref{16}) share the
properties of diffeomorphism covariance and anomaly-freeness. To see 
that the second operator in (\ref{16}) is actually symmetric requires 
a modification of the regularization which does not spoil those 
properties. We will come back to the  modification in \cite{0}.

\subsection{Diffeomorphism covariance}

According to the programme of algebraic quantization proposed in \cite{18}
the solutions to the Euclidean Hamiltonian constraint are diffeomorphism 
invariant
distributions $\psi$ on $\Phi:=\mbox{Cyl}^\infty(\agb)$ such that for all
lapse functions $N$ 
\be \label{20}
\Psi[\hat{H}^E[N]\phi]=0\;\forall\;\phi\in\Phi \;.
\ee
Now take $\phi=f$ to be any function, cylindrical with respect to some 
graph $\gamma$, in the domain of $\hat{H}^E$ then (\ref{20}) amounts to
$\sum_{v\in V(\gamma)}\Psi[\hat{H}^E_v f]=0$ for all $N_v,v\in V(\gamma)$
and therefore we find that we need to satisfy
\be \label{20a}
\Psi[\hat{H}^E_v f_\gamma]=0\;\forall\;\gamma,\;f_\gamma\in
\mbox{Cyl}^3_\gamma(\agb),\; v\in V(\gamma)\;.
\ee
Equation (\ref{20a}) is actually quite unexpected : in section 3.1.3 we 
formulated 
the requirement of diffeomorphism-covariance in terms of the constraint 
at a point and is was far from clear that such an operator actually makes 
sense. Equation (\ref{20a}) is the precise formulation of that concept 
and it is manifestly well-defined.\\ 
Our triangulation adapted to a graph was geared at being diffeomorphism 
covariant for each of its vertices separately meaning that each of
the operators $\hat{H}^E_v$ and even each of the operators 
$\hat{p}_\Delta,\hat{E}(v),\hat{H}^E_\Delta,\hat{h}^E_\Delta$ separately is 
covariantly 
defined. For $\hat{p}_\Delta,\hat{E}(v)$ this follows from the 
manifest covariance of $\hat{p}_e$, namely $\hat{U}(\varphi)\hat{p}_e
\hat{U}(\varphi)^{-1}=\hat{p}_{\varphi(e)}$ for any 
$\varphi\in\mbox{Diff}(\Sigma)$ which in turn is a consequence of the 
fact that $\hat{p}_e$ is defined in terms of $h_e(A)$. For the operators
$\hat{H}^E_\Delta,\hat{h}^E_\Delta$ we argue as follows : let f be 
cylindrical with 
respect to a graph $\gamma$ and let $\varphi\in\mbox{Diff}(\Sigma)$. Then 
the tetrahedra $\Delta(\varphi(\gamma)),\varphi(\Delta(\gamma))$ are 
not necessarily equal to each other. However, the 
graphs $\gamma,\gamma'=\varphi(\gamma)$ are diffeomorphic, therefore 
the topology of the routing of the arcs $a_{ij}(\Delta(\gamma))$ through 
the edges of $\gamma$ and of the arcs $a_{ij}(\Delta(\gamma'))$ through 
the edges of $\gamma'$ as specified in section 3.1.3 is the same since
that prescription depended only on the topology of the graph. 
More specifically, this prescription was shown to be independent of the frame
and therefore coincides for any two vertices $v,v'$ of graphs 
$\gamma,\gamma'$ for which neighbourhoods $U,U'$ exist such that 
$U\cap\gamma$ and $U'\cap\gamma'$ are diffeomorphic.\\
Now just choose
a diffeomorphism $\varphi'$ such that $\varphi'(\gamma')=\gamma'$ and such
that $\varphi'(\Delta(\gamma'))=\varphi(\Delta(\gamma))$ for that specific
$\Delta(\gamma)$ (notice that (\ref{20a}) is a linear combination of 
terms, each of which depends on only one specific $\Delta$ so that we can
adapt $\varphi'$ to $\Delta$). Such a diffeomorphism clearly exists : there 
are diffeomorphisms which leave the image of the graph invariant while 
moving its points and off the graph it can put the arc $a_{ij}(\Delta)$
into any diffeomorphic shape. It follows from these considerations that
$\hat{U}(\varphi')\hat{h}^E_{\Delta(\varphi(\gamma))}\hat{U}(\varphi')^{-1}
=\hat{h}^E_{\varphi(\Delta(\gamma)}$ which is what we wanted to show.\\
Then, obviously, the number $\Psi[\hat{H}^E_v f_\gamma]$
depends only on the diffeomorphism class of the loop assignments
$\alpha_{ij}(\Delta)$ (this was first observed in \cite{12}). 
Therefore, in this diffeomorphism invariant
context, the loops $\alpha_{ij}(\Delta)$ can be chosen as ``small" and the
triangulation as ``fine" as we wish, the value of (\ref{20a}) remains 
invariant and in that sense the continuum limit has already been taken.\\
Diffeomorphism covariance is therefore a sufficient requirement for our
quantum theory to correspond to a continuum theory.\\ 
One might wonder what happens if one actually takes the limit 
and sends $\Delta\to v(\Delta)$. It is easy to see that the result 
vanishes trivially which is not what we want. This happens due to the fact 
that after applying 
$\hat{H}^E(N)$ to a cylindrical function only a finite number of terms 
survive : since a cylindrical function is already determined on smooth 
connections we can, in the limit, actually replace $\hat{H}^E_\Delta$
by an integral over $\Delta$ as in (\ref{10}) but the limit corresponds to
a point which has zero Lebesgue measure. \\
The fact that the limit $\Delta\to v(\Delta)$ is trivial is strange at first
sight because one is used, from the lattice regularization of, say,
$\lambda\phi^4$ theory, that the continuum theory is only recovered if 
we take the lattice spacing to zero, that is, {\em one takes a continuous
cut-off parameter to its continuum value}. In our regularization such a 
parameter simply does not exist and the reason for that is the underlying
diffeomorphism invariance of the theory. This shows that the limit
$\Delta\to v(\Delta)$ is in fact inappropriate.

\subsection{Anomaly-freeness}

Although our assignment is covariant and therefore the continuum limit is
already taken (in the sense explained above) so that it seems that the 
regulator is entirely removed, the operator (\ref{16}) still carries a 
sign of the regularization
procedure : it depends on the diffeomorphism class $[T]$ of the 
triangulation assignment which labels the freedom that we have
in our regularization scheme.
It is therefore not an entirely trivial task to check whether
our operator $\hat{H}^E_T[N]$ is anomaly-free,
meaning that $[\hat{H}^E_T[M],\hat{H}^E_T[N]]f$ vanishes 
for any cylindrical function $f$ and lapse functions $M,N$ when evaluated
on a diffeomorphism invariant state $\psi$. The reason why we do not check
the commutator on $\psi$ immediately is because $\psi$ is a distribution 
\cite{18} and so does not lie in the domain of $(\hat{H}^E(M))^\dagger$. 
Therefore, 
the formal anomaly-freeness condition
$[(\hat{H}^E(N))^\dagger,(\hat{H}^E(M))^\dagger]\psi=0$ has to be 
interpreted in the usual weak sense 
$([(\hat{H}^E(N))^\dagger,(\hat{H}^E(M))^\dagger]\psi)(f)=
\psi([\hat{H}^E(M),\hat{H}^E(N)]f)=0$ for each test function $f$, that
is, every smooth cylindrical function $f\in\Phi$.\\
If the theory is anomaly-free, then, since $\psi$ is a generalized 
eigenvector \cite{18} for the exponentiated
diffeomorphism constraint $\hat{U}(\phi),\phi\in\mbox{Diff}(\Sigma)$ 
with eigenvalue $1$, we expect that in the last equation the argument of
$\psi$ is identically zero if some finite diffeomorphisms can be removed.
This expectation turns out to be precisely correct.\\
We show now that a solution to the anomaly freedom problem is obtained \\
a) for the non-symmetric operator in case we attach the edges 
$a_{ij}(\Delta)$ irrespective of their differentiability at their
endpoints with respect to $s_i(\Delta),s_j(\Delta)$ as prescribed in 
section 3.1.3\\ 
b) for the symmetric operator only {\em if all the loops
$\alpha_{ij}(\Delta)$ are chosen to be kinks with vertex at $v(\Delta)$} ! 
That is, the arc $a_{ij}(\Delta)$ joins the endpoints of 
$e_i(\Delta),e_j(\Delta)$ in at least a $C^1$ fashion (see \cite{0}
for the details of the attachment). An arbitrary attachment of
$a_{ij}(\Delta)$ is insufficient to guarantee anomaly-freeness. \\
Consider for simplicity a graph $\gamma$ which only has one vertex $v$ and 
that it is two-valent, for instance a kink (we ignore for the moment that
functions on such graphs are actually annihilated in order not to 
veil the argument, the problem shows up on higher valent graphs). 
Acting once with the Hamiltonian constraint 
on a function $f$ cylindrical with respect to $\gamma$ we get a function 
cylindrical with respect to a graph $\gamma'$ which contains $\gamma$ and
an additional edge $e$ which intersects $\gamma$ in vertices $v_1,v_2$
and such that the tangents of $e,\gamma$ at these new vertices are, a priori,
linearly independent. Acting once again with the Hamiltonian constraint 
it acts now non-trivially at all three vertices. Therefore, the commutator 
becomes now schematically $[\hat{H}^E(M),\hat{H}^E(N)]f=
(M_v\hat{H}^E_v+M_{v_1}\hat{H}^E_{v_1}+M_{v_2}\hat{H}^E_{v_2})
N_v\hat{H}^E_v f
-(N_v\hat{H}^E_v+N_{v_1}\hat{H}^E_{v_1}+N_{v_2}\hat{H}^E_{v_2})
M_v\hat{H}^E_v f
=(M_{v_1}N_v-N_{v_1}M_v)\hat{H}^E_{v_1}\hat{H}^E_v  f+
(M_{v_2}N_v-N_{v_2}M_v)\hat{H}^E_{v_2}\hat{H}^E_v  f$
which does not manifestly vanish even if 
$[\hat{H}^E_v,\hat{H}^E_{v'}]=0$ for 
$v\not=v'$. One sees that what has to be avoided is that the Hamiltonian 
has non-trivial action at $v_1,v_2$.
The idea is to exploit that the volume 
operator acts trivially on vertices which are such that the tangents of
all edges incident at it lie in a common plane.\\

{\em It is here where the volume operator as defined in the second 
reference of \cite{23}
is selected by the dynamics of the theory while the operator as defined in
\cite{22} has to be rejected (if one follows the approach advertised 
here) !}\\ 

Namely, the volume operator \cite{23} does not annihilate
co-planar vertices. It is very appealing that the requirement of 
anomaly-freeness provides us with a selection rule among the operators 
\cite{22}, \cite{23} which, a priori, from a purely kinematical 
point of
view, are both bona fide quantizations of the classical volume functional.

Now, for the non-symmetric operator there is no problem at all
because by construction of the triangulation assignment the endpoints
of $a_{ij}(\Delta)$ form always {\em new} three-valent vertices of
$\gamma\cup a_{ij}(\Delta)$ but the edges incident at them only have two 
independent tangent directions there, namely those of $e_i(\Delta)$ and 
$a_{ij}(\Delta)$. But for the symmetric operator we need to adjoin
the smooth exceptional edges (see \cite{0}) in at least a $C^1$ 
fashion because all the
segments $a_{ij}(\Delta)$ which intersect the {\em same} edge $e$ of the 
skeleton of the given graph at all, intersect it at the same point and 
in general for higher than valence two the volume operator is non-vanishing
on (not necessarily gauge-invariant) cylindrical functions unless the 
tangents of triples of edges at vertices are linearly dependent.\\ 
In \cite{12,13} the loop assigned does not have the topology of a kink but,
the topology of a triangle. But because these operators do not involve
the volume operator, rather they depend on an operator corresponding
to $\epsilon_{abc}\epsilon_{ijk}E^a_i E^b_j$ which does not vanish
if there are only two independent tangent directions in the problem, 
we expect these operators to be anomalous.
However, if the loop assigned would have the topology of a kink as well
we believe that the anomaly could be removed once these operators are
rigorously defined.
\begin{Theorem} \label{th6}
The Euclidean Hamiltonian operator 
$\hat{H}^E(N)=(\hat{H}^E_\gamma(N),D_\gamma)$ as defined by (\ref{16})
is non-anomalous.
\end{Theorem}
Proof :\\
Let $f$ be a function cylindrical with respect to a graph $\gamma$ and 
let $\Delta(\gamma)$ denote the various tetrahedra attached to it 
in applying $\hat{H}^E(N)$. We may, without loss of generality, assume 
that $f$ is a spin-network function. Clearly the functions
$\hat{H}^E_{\Delta(\gamma)}f,\hat{h}^E_{\Delta(\gamma)}f$ 
depend on the 
graph $\gamma\cup\Delta(\gamma)$. We are being very explicit here in the 
dependence of the tetrahedra on the graph because this will be essential 
in what follows. We may assume that all the tetrahedron projectors 
$\hat{p}_\Delta$ are non-vanishing on $f$, that is, the dependence of 
$f$ on all the edges of $\gamma$ is non-trivial. Then $\hat{E}(v)f=
E(v,\gamma)f=n(n-1)(n-2)/6f$ where $n$ is the valence of $v$ in the graph 
$\gamma$. Now consider $\Delta'$ with $v':=v(\Delta')\not=v(\Delta)=:v$. Then
$\hat{p}_{\Delta'}\hat{H}^E_\Delta f=\hat{H}^E_\Delta f$ since the 
segments of edges of $\gamma$ which may have been removed in 
$\hat{H}^E_\Delta f$ are at a vertex different from the vertex 
$v(\Delta')$. It follows that $\hat{E}(v')\hat{H}^E_\Delta f=
E(v',\gamma)\hat{H}^E_\Delta f$. If, however, $v=v'$ then it is possible
that $\hat{p}_{\Delta'}g=0$ where $g$ is a spin-network 
state appearing in the decomposition of $\hat{H}^E_\Delta f$. Now, for 
the symmetric operator, since we project with $\hat{p}_\Delta$ before and 
after applying $\hat{h}^E_\Delta$, such $g$ are automatically removed
so that either $g=0$ or $\hat{E}(v')g=E(v,\gamma)g$ again. For the 
non-symmetric operator we get instead $\hat{E}(v')g=E(v,\gamma,g)g$.
With this preparation 
we compute for the non-symmetric operator (it is understood that only terms 
with  
$\hat{p}_{\Delta(\gamma\cup\Delta(\gamma))}\hat{H}^E_{\Delta(\gamma)}\not=0$ 
are kept)
\ba \label{21} 
& &\hat{H}^E[M]\hat{H}^E[N] f =\sum_{v\in V(\gamma)} \frac{N_v}{E(v,\gamma)}
\sum_{v(\Delta(\gamma))=v} \hat{H}^E[M]\hat{H}^E_{\Delta(\gamma)}f\nonumber\\
&=& \sum_{v\in V(\gamma)} \frac{N_v}{E(v,\gamma)}
\sum_{v(\Delta(\gamma))=v}\;\;\sum_{v'\in V(\gamma\cup\Delta(\gamma)}
M_{v'} 
\sum_{v(\Delta(\gamma\cup\Delta(\gamma)))=v'}
\hat{H}^E_{\Delta(\gamma\cup\Delta(\gamma))}
\frac{1}{\hat{E}(v')}\hat{H}^E_{\Delta(\gamma)}f\nonumber\\
&&
\ea
and similar for the symmetric operator,
where the essential step has been the last one where all the contributions
from $V(\gamma\cup\Delta(\gamma))-V(\gamma)$, where $v(\Delta(\gamma))\in 
V(\gamma)$ 
have been removed due to the fact that all these vertices are co-planar
(for the symmetric operator this holds because of 
the kink property of those vertices as explained in Lemma 3.1 and
the subsequent Collorary).
We now compute (\ref{21}) again with the roles of $M,N$ interchanged,
subtract from (\ref{21}) and obtain 
\ba \label{22}
&&[\hat{H}^E[M],\hat{H}^E[N]] f\nonumber\\
&&=\sum_{v,v'\in V(\gamma)} (M_{v'} N_v -M_v N_{v'})\frac{1}{E(v,\gamma)}
\sum_{v(\Delta(\gamma))=v,v(\Delta(\gamma\cup\Delta(\gamma)))=v'}
\hat{H}^E_{\Delta(\gamma\cup\Delta(\gamma))}\frac{1}{\hat{E}(v')}
\hat{H}^E_{\Delta(\gamma)}f
\nonumber\\
&&=\sum_{v,v'\in V(\gamma),v<v'} \frac{M_{v'} N_v -M_v N_{v'}}{E(v)E(v')}
\times\nonumber\\
&&\times\sum_{v(\Delta(\gamma))=v(\Delta(\gamma\cup\Delta'(\gamma)))=v,
v(\Delta'(\gamma))=v(\Delta'(\gamma\cup\Delta(\gamma)))=v'}
[\hat{H}^E_{\Delta'(\gamma\cup\Delta(\gamma))}\hat{H}^E_{\Delta(\gamma)}
-\hat{H}^E_{\Delta(\gamma\cup\Delta'(\gamma))}\hat{H}^E_{\Delta'(\gamma)}]f
\nonumber\\
&&
\ea
where in the second step the notation $v<v'$ assumes that we have ordered 
the vertices of $\gamma$ somehow which allowed us to write a sum 
over unordered pairs of vertices. Also we used that the 
antisymmetric product of lapse functions vanishes at equal vertices which 
was crucial in replacing $\hat{E}(v')$ by $E(v',\gamma)=:E(v')$ so that 
we may imagine to absorb this number into the lapses, $N_v/E(v)\to N_v$ (as 
explained above, this happens for the symmetric operator already before 
taking the commutator). Formula (\ref{22}) is valid for the symmetric 
operator as well if we just replace $\hat{H}^E_\Delta$ by 
$\hat{h}^E_\Delta$.\\
It is far from obvious whether (\ref{22}) vanishes or not. Indeed, for 
genuine $\gamma$ and genuine choices of loop assignments depending on the 
graph, (\ref{22}) is a non-vanishing cylindrical function of positive 
$L_2$ norm. We now evaluate a diffeomorphism invariant state on 
(\ref{22}). We can take the antisymmetric product of the lapse functions 
evaluated at different vertices out of the integral over $\agb$ and it 
will be sufficient to show that 
\be \label{22a}
\psi[(\hat{H}^E_{\Delta'(\gamma\cup\Delta(\gamma))}\hat{H}^E_{\Delta(\gamma)}
-\hat{H}^E_{\Delta(\gamma\cup\Delta'(\gamma))}\hat{H}^E_{\Delta'(\gamma)})f]=0
\ee
for each choice of 
$v(\Delta(\gamma))=v(\Delta(\gamma\cup\Delta'(\gamma)))=v,
v(\Delta'(\gamma))=v(\Delta'(\gamma\cup\Delta(\gamma)))=v'$
separately. To see this, notice first that the members of the first pair of 
tetrahedra given by
$(\Delta(\gamma),\Delta(\gamma\cup\Delta'(\gamma)))$ 
as well as the member of the second pair of tetrahedra given by 
$(\Delta'(\gamma),\Delta'(\gamma\cup\Delta(\gamma)))$ 
are diffeomorphic. This follows immediately from the fact that 
$v\not=v'$ so that there are disjoint neighbourhoods $U$ of $v$ and $U'$ of 
$v'$ where $U$ and $U'$ respectively contain both members of the first 
and second pair of tetrahedra respectively. Let 
$\varphi\in\mbox{Diff}(\Sigma)$ be chosen such that $\gamma\cap U'$ is
left invariant and but that $\Delta(\gamma\cup\Delta'(\gamma))
=\varphi(\Delta(\gamma))$. Likewise, choose $\varphi'\in\mbox{Diff}(\Sigma)$ such 
that $\gamma\cap U$ is left invariant but that 
$\Delta'(\gamma\cup\Delta(\gamma))=\varphi'(\Delta'(\gamma))$. 
Then, using diffeomorphism invariance of $\psi$ we 
find that the left hand side of the (\ref{22a}) becomes
\ba \label{22b}
&&\psi[(\hat{U}(\varphi')\hat{H}^E_{\Delta'(\gamma)}\hat{H}^E_{\Delta(\gamma)}
-\hat{U}(\varphi)\hat{H}^E_{\Delta(\gamma)}\hat{H}^E_{\Delta'(\gamma)})f]
\nonumber\\
&=&\psi[(\hat{U}(\varphi')\hat{H}^E_{\Delta'(\gamma)}\hat{H}^E_{\Delta(\gamma)}
\hat{U}(\varphi^{-1})
-\hat{U}(\varphi)\hat{H}^E_{\Delta(\gamma)}\hat{H}^E_{\Delta'(\gamma)}
\hat{U}(\varphi')^{-1})f]\nonumber\\
&=&\psi[[\hat{H}^E_{\Delta'(\gamma)},\hat{H}^E_{\Delta(\gamma)}]f]\;.
\ea
That is, we were able to ``match" the tetrahedra using diffeomorphism 
invariance. In the second equality we used the invariance 
$\hat{U}(\varphi)f=f \Rightarrow f=\hat{U}^{-1}(\varphi)f$ and similar for 
$\varphi'$ in order to write the commutator in such a way that it 
becomes manifestly antisymmetric if we replace $\hat{H}^E_\Delta\to
\hat{h}^E_\Delta$.\\ 
Now we just need to use Collorary 3.1 of \cite{0} (the non-symmetric operator 
is treated in a comment after this collorary) to see 
that the commutator in (\ref{22b}) vanishes identically because the part 
$\hat{V}_v$ of the volume operator involved in 
$\hat{H}^E_{\Delta(\gamma)}$
does not act on the holonomies along edges incident at $v'$ involved in 
$\hat{H}^E_{\Delta'(\gamma)}$ and vice versa since the vertices $v,v'$ 
are different. That completes the proof of anomaly-freeness.\\
$\Box$\\
\\
A couple of remarks are in order :\\
$\bullet$ The classical constraint algebra is given by
$\{H^E[M],H^E[N]\}=\int_\Sigma (M N_{,a}-N M_{,a})q^{ab} V_b$, where 
$V_a$ is the classical diffeomorphism constraint and $q^{ab}$ is the 
inverse metric tensor. Naively, one would expect that the quantum version
of that would be 
\be \label{22c}
\frac{1}{i\hbar}[\hat{H}^E[M],\hat{H}^E[N]]=
\int_\Sigma (M N_{,a}-N M_{,a})*\hat{q}^{ab} \hat{V}_b* 
\ee
where the $*...*$ is to indicate that the operators that appear have to be
regularized and ordered appropriately. The immediate problem with (\ref{22c})
as, widely discussed in the literature, is that the constraint algebra is
not a proper algebra, the structure functions depend on the canonical
variables through $q^{ab}$ and if in the quantum theory, following the 
Dirac approach, the generator of the
diffeomorphism constraint does not appear to the right of $\hat{q}^{ab}$ 
in (\ref{22c}) then one would not expect a genuine diffeomorphism invariant 
state 
to be annihilated by the commutator of two Hamiltonian constraints : One says
that the constraint algebra has an anomaly and that the quantum theory 
does not correspond to the classical theory because any element of the kernel
of both the diffeomorphism constraint and the Hamiltonian constraint would
have to satisfy the additional requirement that the right hand side of
(\ref{22c}) annihilates it which reduces the number of degrees of freedom
more than the classical theory would do.\\
One can then ask the question whether there exists a 
consistent quantization (regularization and factor ordering) of 
$\hat{H}^E(N)$ such that there is no anomaly. In \cite{Kuchar} the 
authors investigate a wide class of finite-dimensional theories (gauge 
systems) with 
a Hamiltonian quadratic in the momenta and a constraint algebra which 
mimics (\ref{22c}) algebraically in the sense that $V_b$ is replaced by 
the generator of a gauge transformation and $q^{ab}$ is replaced by some 
non-constant function of the canonical configuration coordinates. The 
authors find that a consistent
quantization can be obtained but {\em never in such a way that the 
Hamiltonian constraint is a symmetric operator}. The intuitive reason for 
this is clear : since the generator $\hat{V}_a$ of the unitary 
representation of the Diffeomorphism group has to be self-adjoint and 
since the metric tensor should be at least symmetric as well, then
the left hand side of (\ref{22c}) should appear in the symmetric ordering
$\hat{q}^{ab}\hat{V}_{b}+\hat{V}_a\hat{q}^{ab}$ (or an even more 
complicated symmetric operator) whenever $\hat{H}^E(N)$ is a symmetric 
operator.
Now, in one of its versions, we actually {\em did} order $\hat{H}^E(N)$ 
symmetrically ! Why, then,
is that not in contradiction to the arguments given in \cite{Kuchar} 
and references therein ?\\
As is often the case with ``no-go theorems", one needs to carefully check 
the assumptions. The assumptions underlying the considerations in
\cite{Kuchar}, when applied to our case, contain at least the following
list (we do not list 
the assumption that one only has a finite number of degrees of freedom 
since we do not want to blame the failure of the theorem on that) :\\ 
1) The operators $\hat{H}^E(N),\hat{q}^{ab}(x),\hat{V}_b(x)$ 
can be regulated and densely defined on $\cal H$.\\
2) The classical Hamiltonian constraint is a bilinear form in the canonical 
momenta.\\ 
3) The classical Diffeomorphism constraint is a linear form in the 
canonical momenta.\\
Only assumption 3) is satisfied here, the rest is violated : \\
1) As was shown in \cite{8} the representation of (one-parameter subgroups 
of) the diffeomorphism group
on $\cal H$ is not strongly continuous. Therefore, by Stone's theorem,
there does not exist a self-adjoint generator. Also a operator corresponding
to $q^{ab}(x)$ is entirely meaningless in our representation : one could 
write it as $q^{ab}=E^a_i E^b_i/\det(q)$, however, both nominator and 
denominator are meaningless \cite{18}.\\
2) The dependence of $H^E$ on the momenta is through $\{A_a,V\}$ which is 
not even an analytic function of $E^a_i$.\\
We conclude that the considerations of \cite{Kuchar} are not applicable in
to our case.\\
How then can we even define a theory to be anomaly-free if we cannot define
the quantizations of the generators of the symmetries of the classical 
theory ? The answer is that the quantization of a classical theory does not 
force us to make 
sense out of every classical function as an operator. The most general 
question we can ask is the following : \\
a) Can we make sense out of the right hand side of (\ref{22c}), that is,
can we make sense out of 
\be \label{22d}
\widehat{\int_\Sigma (M N_{,a}-N M_{,a})q^{ab} V_b}\; ? 
\ee
The answer is, trivially, in the affirmative : up to higher order in
$\hbar$ the commutator $[\hat{H}^E(M),\hat{H}^E(N)]/(i\hbar)]$ reproduces 
$\int_\Sigma (M N_{,a}-N M_{,a})q^{ab} V_b$ so that we can just
{\em define} (\ref{22d}) by that commutator because by construction it
is well-defined.\\
b) The fact that the classical Poisson bracket vanishes on the constraint
surface of the phase space defined by the diffeomorphism constraint
translates in the quantum theory into the requirement that (\ref{22d})
should vanish on diffeomorphism invariant states and this we checked to 
be the case.\\
To conclude, in our case the question of whether a factor ordering can be 
found such that $\hat{V}_a$ stands to the right in (\ref{22d}) cannot 
even be asked. Therefore the quantum constraint algebra cannot have any 
close analogy with the classical constraint algebra. Given the fact that 
we can only define an operator corresponding to finite diffeomorphisms
the structure of the constraint algebra that we {\em do} expect is 
precisely the one displayed in (\ref{22b}), namely that
the commutator on a state $f$ is a sum of terms of the form 
$\hat{U}(\varphi)\hat{A}\hat{B}f-\hat{U}(\varphi')\hat{B}\hat{A}f$
where $[\hat{A},\hat{B}]f=0$ and $\hat{A},\hat{B}$ are symmetric 
operators. The diffeomorphisms $\varphi,\varphi'\in\mbox{Diff}(\Sigma)$
are highly ambiguous since their action has only been specified on a 
finite graph induced from the one underlying $f$, in particular 
$\hat{U}(\varphi)f=\hat{U}(\varphi')f=f$. 
Using this ambiguity we can  
write the right hand side of the commutator $[\hat{H}^E(M),\hat{H}^E(N)]f$
again in manifestly anti-symmetric form 
$\hat{U}(\varphi)\hat{A}\hat{B}f\hat{U}(\varphi')^{-1}
-\hat{U}(\varphi')\hat{B}\hat{A}\hat{U}(\varphi)^{-1}f$
as in (\ref{22b}) so that there are no 
factor-ordering contradictions of the kind discussed in \cite{Kuchar} at 
all !\\
$\bullet$ It will turn out that we do not need to apply any group 
averaging \cite{18} with respect  
to the non-symmetric Hamiltonian constraint operator but rather can 
compute the solutions by direct methods. This is important because if the 
constraint operator is not symmetric then the group averaging method 
cannot be immediately applied. However, for the symmetric operator, 
to which we can apply the method, direct methods are not possible to 
apply and so group averaging becomes important. Expectedly, in the 
diffeomorphism invariant context it is true that the constraint algebra 
becomes Abelian which makes group averaging especially attractive
(see \cite{0}).

\section{Generator of the Wick rotation transform}

As explained in section 3, since classically the generator is given by
$C=(\pi/2) K$, almost no work is needed to do. We just notice that 
$K=-\{V/\kappa,H^E[1]\}$ and now simply define
\be \label{23}
\hat{K}:=-\frac{1}{i\ell_p^2}[\hat{V},\hat{H}^E[1]]\mbox{ and }
\hat{C}:=\frac{\pi}{2}\hat{K} \;.\ee
Notice that $\hat{V},\hat{H}^E$ have both dense domain and range in
$\mbox{Cyl}^3(\agb)$. Therefore, $\hat{K}$ has 
also dense domain and range in $\mbox{Cyl}^3(\agb)$. Moreover, consider the 
case that we are using the symmetric version of $\hat{H}^E(N)$.
Then $\hat{K}$ is also symmetric and 
since $\hat{H}^E[1]$ is an imaginary-valued
operator, $\hat{K}$ and $\hat{V}$ are both real-valued operators.
We now choose complex conjugation as the conjugation in von Neumann's 
theorem (see \cite{0}) and see that both $\hat{V},\hat{K}$ have then 
self-adjoint extensions.\\
Remark :\\
Notice that the method of getting $\hat{K}$ from 
$\hat{V},\hat{H}^E[1]$ simply through the commutator of these two operators
is not possible to apply using any of the other operators corresponding to 
$H^E[1]$ as defined so far in the literature : the classical identity 
$K=-\{V,H^E[1]\}$ holds only for
$H^E$ given by (\ref{7}), it does not hold either for $\sqrt{\det(q)}H^E$
\cite{26,27} or for $\sqrt{\sqrt{\det(q)}H^E}$ \cite{12,13} and thus for 
those approaches the operator corresponding to the generator of the 
Wick rotation transform is far from easy to define.\\
For later purposes we wish to derive a more explicit expression for
$\hat{K}$. Let $f$ be a function cylindrical with respect to a graph
$\gamma$. Then we have 
\be \label{24}
\hat{K}f =-\frac{1}{i\ell_p^2}\sum_{v,v'\in V(\gamma)}
[\hat{V}_{v'},\hat{H}^E_v]f 
= -\frac{1}{i\ell_p^2}\sum_{v\in V(\gamma)}
[\hat{V}_{v},\hat{H}^E_v]f=:\sum_{v\in V(\gamma)} \hat{K}_v f
\ee
where in the first step we again used the fact that the volume operator 
does not see the vertices $V(\gamma\cup\Delta(\gamma))-V(\gamma)$ and in the 
second step we exploited that $\hat{V}_v$ only acts on the edges incident
at $v$ so that the commutator with $\hat{H}^E_{v'}$, which contains only
holonomies of edges incident at $v'$, vanishes if $v'\not=v$. So, each term
$\hat{K}_v$ in the sum over vertices contains only the part $\hat{V}_v$ of 
the full volume operator.

\section{The Wheeler-DeWitt constraint operator}

As outlined in section 3, we will now first derive a regulated operator
corresponding to the expression 
$\mbox{tr}([K_a,K_b][E^a,E^b])/\sqrt{\det(q)}$ which is
consistently defined, whose action is diffeomorphism covariant
and, in case we are 
dealing with a symmetric $\hat{H}^E(N)$, is symmetric as well. 
In the symmetric case the idea is then to treat it as a perturbation of some 
self-adjoint extension 
of $\hat{H}^E$ in the expression of $\hat{H}$ and to try to invoke the 
Kato-Rellich 
theorem to conclude that it is self-adjoint on the domain of $\hat{H}^E$
(see \cite{0}).
Finally we will prove that both versions of the operator are 
anomaly-free.

\subsection{Regularization}

We will not repeat all the arguments here as the derivation is completely
analogous to the one for $\hat{H}^E$.\\
We begin with the classical expression for the ``kinetic term"
\be \label{25}
T[N]:=8\int_\Sigma d^3x 
\frac{N}{\kappa^3}\epsilon^{abc}\mbox{tr}(\{A_a,K\}\{A_b,K\}\{A_c,V\}) \ee
and introduce the same triangulation $T$ as used for $H^E$ to regulate the 
integral with the result 
\be \label{26}
T[N]_T:=-\frac{8}{3\kappa^3}\sum_{\Delta\in T} N(v(\Delta))
\epsilon^{ijk}\mbox{tr}(h_{s_i(\Delta)}\{h_{s_i(\Delta)}^{-1},K\}
h_{s_j(\Delta)}\{h_{s_j(\Delta)}^{-1},K\}h_{s_k(\Delta)}
\{h_{s_k(\Delta)}^{-1},V\}). 
\ee
Here, $N$ is again the lapse function divided by $\kappa$.
Each term in the sum (\ref{26}) labelled by a tetrahedron 
$\Delta$ tends to
$8/\kappa^3\int_\Delta N
\mbox{tr}(\{A,K\}\wedge\{A,K\}\wedge\{A,V\})$ as we shrink 
the tetrahedra to their basepoints, that is, we get the correct continuum 
limit.\\
The quantization of this expression now merely consists in adapting
$T$ to a graph as outlined for the regularization of 
$\hat{H}^E$ and in replacing Poisson brackets by commutators and functions
by operators. The result, when applied to a function cylindrical with 
respect to a graph $\gamma$, reads
\ba \label{27}
&&\hat{T}[N]f =-\frac{64}{3(i\ell_p^2)^3}\sum_{v\in V(\gamma)} N_v
\frac{1}{E(v)}\sum_{v(\Delta)=v}
\epsilon^{ijk}\times\nonumber\\
& & \times \mbox{tr}(h_{s_i(\Delta)}[h_{s_i(\Delta)}^{-1},K_v]
h_{s_j(\Delta)}[h_{s_j(\Delta)}^{-1},K_v]h_{s_k(\Delta)}
[h_{s_k(\Delta)}^{-1},V_v]) f\nonumber\\
&&=:\sum_{v\in 
V(\gamma)}\frac{N_v}{E(v)}\sum_{v(\Delta)=v}\hat{T}_\Delta f
=:\sum_{v\in V(\gamma)}\frac{N_v}{E(v)}\hat{T}_v f \;. \ea
In order to arrive at this expression we use the same arguments as before
to conclude that only tetrahedra intersecting a vertex of $\gamma$ give a 
contribution and we realize that all the commutators with holonomies 
along edges incident at $v$ are non-vanishing only for the parts $\hat{K}_v$
and $\hat{V}_v$ of $\hat{V}$ and $\hat{K}$ respectively. To prove the 
latter we merely need to observe the following :\\
Let $s$ be any segment of an edge of $\gamma$ incident at a vertex $v$ then 
$g:=[h_s,\hat{K}]f=\sum_{v'\in V(\gamma)} h_s
\hat{K}_v f -\sum_{v\in V(\gamma\cup\{s\})} \hat{K}_v h_s f=[h_s,K_v]f$ 
because of the already familiar argument that $\hat{K}_{v'}$ only involves 
the volume operator at $v'$ which commutes with $h_s$ unless $v=v'$ and we
also used $s\cup\gamma=\gamma$. Next we 
have already seen that
$h:=h_{s_j(\Delta)}[h_{s_j(\Delta)}^{-1},\hat{V}]g
=h_{s_j(\Delta)}[h_{s_j(\Delta)}^{-1},\hat{V}_v]g$
which is a cylindrical function with respect to $\gamma\cup\Delta(\gamma)$.
Now, since for $v\in\gamma$ and $v'\in V(\gamma\cup 
\Delta(\gamma))-V(\gamma)$ we have 
trivially $[h_{s_i(\Delta)}^{-1},\hat{K}_{v'}]h=0$ it follows that 
$h_{s_i(\Delta)}[h_{s_i(\Delta)}^{-1},\hat{K}]h=
h_{s_i(\Delta)}[h_{s_i(\Delta)}^{-1},\hat{K}_v]h$ as claimed.
This furnishes the regularization part.

\subsection{Cylindrical Consistency}

Notice that in (\ref{27}) there appear factors of 
$h_{s_i(\Delta)},\Delta=\Delta(\gamma)$
which are holonomies along segments of edges of the {\em original} graph
$\gamma$. Now, according to the definition of $\hat{K}$ one is supposed to 
adapt the 
triangulation associated with $\hat{K}$ (which is of course the same as 
the one associated with $\hat{H}^E$) to the graph 
$\gamma\cup s_i(\Delta(\gamma))$ when acting with $\hat{K}$ on 
$h_{e_i(\Delta(\gamma))}f_\gamma$ for instance. However, $\hat{K}$ is 
consistently defined up to a diffeomorphism and
$\gamma\cup e_i(\Delta(\gamma))=\gamma$ so that 
(\ref{27}) is cylindrically consistently defined up to a diffeomorphism,
provided we introduce again the projectors $\hat{p}_\Delta$. So we redefine
\be \label{27a}
\hat{T}_v\;\to\; \hat{T}_v:=\sum_{v(\Delta)=\Delta}\hat{T}_\Delta
\frac{\hat{p}_\Delta}{\hat{E}(v)}
\ee
for the non-symmetric version of $\hat{H}(N)$ while for the symmetric 
version we refer to (3.5), (3.6) and (3.7) of \cite{0}.

\subsection{Diffeomorphism-covariance}

Diffeomorphism covariance is immediate because $\hat{T}$ is covariantly
defined if and only $\hat{H}^E$ is. There are no loop assignments different
from the ones made for $\hat{H}^E$ involved in the definition of $\hat{T}$.

\subsection{Anomaly-freeness}

\begin{Theorem}
The complete (symmetric or non-symmetric) Lorentzian Wheeler-Dewitt operator
\be \label{30}
\hat{H}_\gamma[N]:=\hat{T}_\gamma[N]-\hat{H}^E_\gamma[N]=\sum_{v\in 
V(\gamma)} N_v[\hat{T}_v-\hat{H}^E_v]
\ee 
is anomaly-free. 
\end{Theorem}
Proof :\\
To see this we use the fact that neither $\hat{H}^E$ 
nor $\hat{T}$ act at the additional vertices introduced by acting 
with $\hat{H}^E,\hat{T}$ (in the symmetric case this requires Lemmata 
3.1 and 3.2
and find that for a spin-network function $f$ cylindrical with respect to 
$\gamma$ we have \be \label{31}
[\hat{H}[M],\hat{H}[N]]f=\sum_{v\not=v'\in V(\gamma)}
\frac{M_v N_{v'}-M_{v'} N_v}{E(v)E(v')}
[\frac{1}{2}\{[\hat{T}_v,\hat{T}_{v'}]+[\hat{H}^E_v,\hat{H}^E_{v'}]\}
-[\hat{T}_v,\hat{H}^E_{v'}]]f 
\ee
that is, again the vertices of every $V(\gamma\cup\Delta(\gamma))-V(\gamma)$ 
were irrelevant.
In (\ref{31}) it is understood that the operators $\hat{T}_v,\hat{H}^E_v$
still depend on the graph on which the function depends that they are 
acting on in complete analogy with the considerations made in (\ref{22}).
Now we see that for $v=v'$ the factor consisting of the antisymmetric
product of the lapse functions vanishes (which was crucial in replacing
$\hat{E}(v')$ by $E(v')=E(v',\gamma)$) while for $v\not=v'$ 
the commutators vanishes when evaluated on a diffeomorphism invariant 
state as in (\ref{22b}) after getting rid of some diffeomorphism 
operators, again 
because all operators involved in the commutators only contain the parts 
$\hat{V}_v, \hat{V}_{v'}$ of the volume operator which do not act on 
edges incident at different vertices.\\
Let us be more specific. Notice that the action of the operator 
$\hat{T}_{\tilde{\Delta}}$, $v$ a vertex of $\gamma$ such that 
$v(\tilde{\Delta})=v,\tilde{\Delta}=\tilde{\Delta}(\gamma)$, on $f$ looks in 
full detail like this ($c$ is a universal constant)
\ba
\hat{T}_{\tilde{\Delta}}&=&\frac{c}{E(v,\gamma)^2}\sum_{v(\Delta(\gamma))=v}
\sum_{v(\Delta'(\gamma\cup\Delta(\gamma)))=v}\epsilon^{ijk}\times\nonumber\\
&&\times\mbox{tr}(h_{s_i(\tilde{\Delta})}[h_{s_i(\tilde{\Delta})}^{-1},
[\hat{V}_v,\epsilon^{lmn}\mbox{tr}(h_{\alpha_{lm}}(\Delta')h_{s_n(\Delta')}
[h_{s_n(\Delta')}^{-1},\hat{V}_v])]]\frac{\hat{p}_{\Delta'}}{\hat{E}(v)}
\times\nonumber\\
&& \times
h_{s_j(\tilde{\Delta})}[h_{s_i(\tilde{\Delta})}^{-1},
[\hat{V}_v,\epsilon^{pqr}\mbox{tr}(h_{\alpha_{pq}}(\Delta')h_{s_r(\Delta)}
[h_{s_r(\Delta)}^{-1},\hat{V}_v])]]
h_{s_k(\tilde{\Delta})}[h_{s_i(\tilde{\Delta})}^{-1},\hat{V}_v])f \;.
\nonumber
\ea
While this is a complicated expression, what counts is that the 
tetrahedron projectors $\hat{p}_{\Delta'(\gamma\cup\Delta(\gamma))}$ can be 
replaced by $\hat{p}_{\Delta'(\gamma)}$, that is, a projector with 
respect to the {\em original} graph $\gamma$ and therefore also
$\hat{E}(v,\gamma\cup\Delta(\gamma))$ can be replaced by $\hat{E}(v,\gamma)$. 
The same is true for the projectors that are involved in the operator
$\hat{T}_{\tilde{\Delta}'}$ or $\hat{H}^E_{\tilde{\Delta}'}$ when applying 
it to $\hat{T}_{\tilde{\Delta}}f$
or $\hat{H}^E_{\tilde{\Delta}}f$ where $v(\tilde{\Delta}')=v'\not=v$.
The associated tetrahedra involved depend on a nested sequence of graphs,
that is, we have terms involving tetrahedra depending on
$\gamma\subset\gamma\cup\Delta(\gamma)\subset\gamma\cup\Delta(\gamma)
\cup\Delta'(\gamma\cup\Delta(\gamma))
\subset\gamma\cup\Delta(\gamma)\cup\Delta'(\gamma\cup\Delta(\gamma))
\cup\Delta^{\prime\prime}
(\gamma\cup\Delta(\gamma)\cup\Delta'(\gamma\cup\Delta(\gamma)))$.
However, just as in the case of computing the commutator between two
Euclidean Hamiltonian constraints, we see that we get a sum of 
terms each of which looks like $\hat{U}(\varphi)\hat{A}\hat{B}-
\hat{U}(\varphi')\hat{B}\hat{A}$. This happens for each pair 
of values that the pair of involved operators 
($\hat{p}_{\tilde{\Delta}}{\hat{E}(v)},
\hat{p}_{\tilde{\Delta}'}{\hat{E}(v')}$ may take because they act at 
different vertices and so it is irrelevant which acts first, they 
commute. This, together with the fact that the volume operator at one 
vertex does not act on the tetrahedra incident at another vertex 
shows that $\hat{A},\hat{B}$ commute upon removing the diffeomorphisms 
$\varphi,\varphi'$.\\ 
This furnishes the proof.\\
$\Box$\\
In the companion paper \cite{0} we will compute the complete solution to 
both the non-symmetric Euclidean and Lorentzian as well as the 
Diffeomorphism constraint. Furthermore we define the symmetric version
of the Lorentzian operator show that it is diffeomorphism-covariant
and non-anomalous as well, however, we do not have the complete kernel in 
this case. Also we will report on the status of the Wick rotation transform
in the light of these two articles.\\
Finally, we will conclude with an interpretation of the results obtained, 
in particular, why the series consisting of these two papers was named
``Quantum Spin Dynamics (QSD)".\\ 
\\
\\
{\large Acknowledgments}\\
\\
I am indebted to Jurek Lewandowski who taught me,
in the framework of his version of the Euclidean Hamiltonian constraint
derived in collaboration with Abhay Ashtekar,
a) how to combine the  requirement of cylindrical consistency with the 
diffeomorphism covariance of the loop assignment of the regularization and 
b) how to prescribe the topology of the routing of the arcs in a 
diffeomorphism invariant fashion.\\
This research project was 
supported in part by DOE-Grant DE-FG02-94ER25228 to Harvard University.\\

\end{document}